\documentclass[Afour,sagev,times]{sagej}

\usepackage{moreverb,url}

\usepackage[colorlinks,bookmarksopen,bookmarksnumbered,citecolor=red,urlcolor=red]{hyperref}

\newcommand\BibTeX{{\rmfamily B\kern-.05em \textsc{i\kern-.025em b}\kern-.08em
T\kern-.1667em\lower.7ex\hbox{E}\kern-.125emX}}

\def\volumeyear{2016}

\usepackage{graphicx}% Include figure files
\usepackage{bm, amssymb, color}% bold math
\usepackage{hyperref}% add hypertext capabilities
\usepackage{natbib}
\usepackage{booktabs,tabulary}
\usepackage{multirow}
\usepackage{amsmath,soul}
\usepackage{caption}
\usepackage{subfigure}
\usepackage{float}
\begin{document}

\runninghead{Extraordinary Disordered Hyperuniform Multifunctional Composites}

\title{Extraordinary Disordered Hyperuniform Multifunctional Composites}

\author{Salvatore Torquato \affilnum{1} }

\affiliation{\affilnum{1}Department of Chemistry, Department of Physics, Princeton Institute of Materials, and
Program in Applied and Computational Mathematics, Princeton University, Princeton, New Jersey 08544, USA}

\email{torquato@princeton.edu}

\begin{abstract}
A variety of performance demands are being placed on material
systems, including desirable  mechanical, thermal, electrical, optical,
acoustic and flow properties. The purpose  of the present article is to review the emerging field
of disordered hyperuniform composites and their novel multifunctional characteristics.
Disordered hyperuniform media are exotic amorphous states of matter
that are characterized by an anomalous suppression of large-scale volume-fraction
fluctuations compared to those in ``garden-variety" disordered materials.
Such unusual composites can have advantages over their periodic counterparts, 
such as unique or nearly optimal, direction-independent
physical properties and robustness against defects. It will be shown
that disordered hyperuniform composites and porous media
can be endowed with a broad spectrum of extraordinary physical properties, including photonic, phononic, transport, chemical and mechanical characteristics  that are only beginning to be discovered.
\end{abstract}


\maketitle

\section{Introduction}

Increasingly, a variety of performance demands are being placed on material
systems.  In aerospace and space applications these requirements include
lightweight component structures that have desirable  mechanical, thermal, electrical, optical,
acoustic and flow properties. Structural components should be able
to carry mechanical loads while having other beneficial performance characteristics. Desirable thermal properties include high thermal conductivity
to dissipate heat and thermal expansion characteristics  that match the attached components.
In the case of porous cellular solids, heat dissipation can be improved by forced
convection through the material, but in these instances the fluid permeability of the
porous material must be large enough to minimize power requirements for convection. Desirable
optical and acoustic properties include materials that can control the propagation of
light and sound waves through them. It is difficult to find single homogeneous
materials that possess these multifunctional characteristics.

\begin{figure}[H]
\centerline{\includegraphics[  width=2.5in,keepaspectratio,clip=]{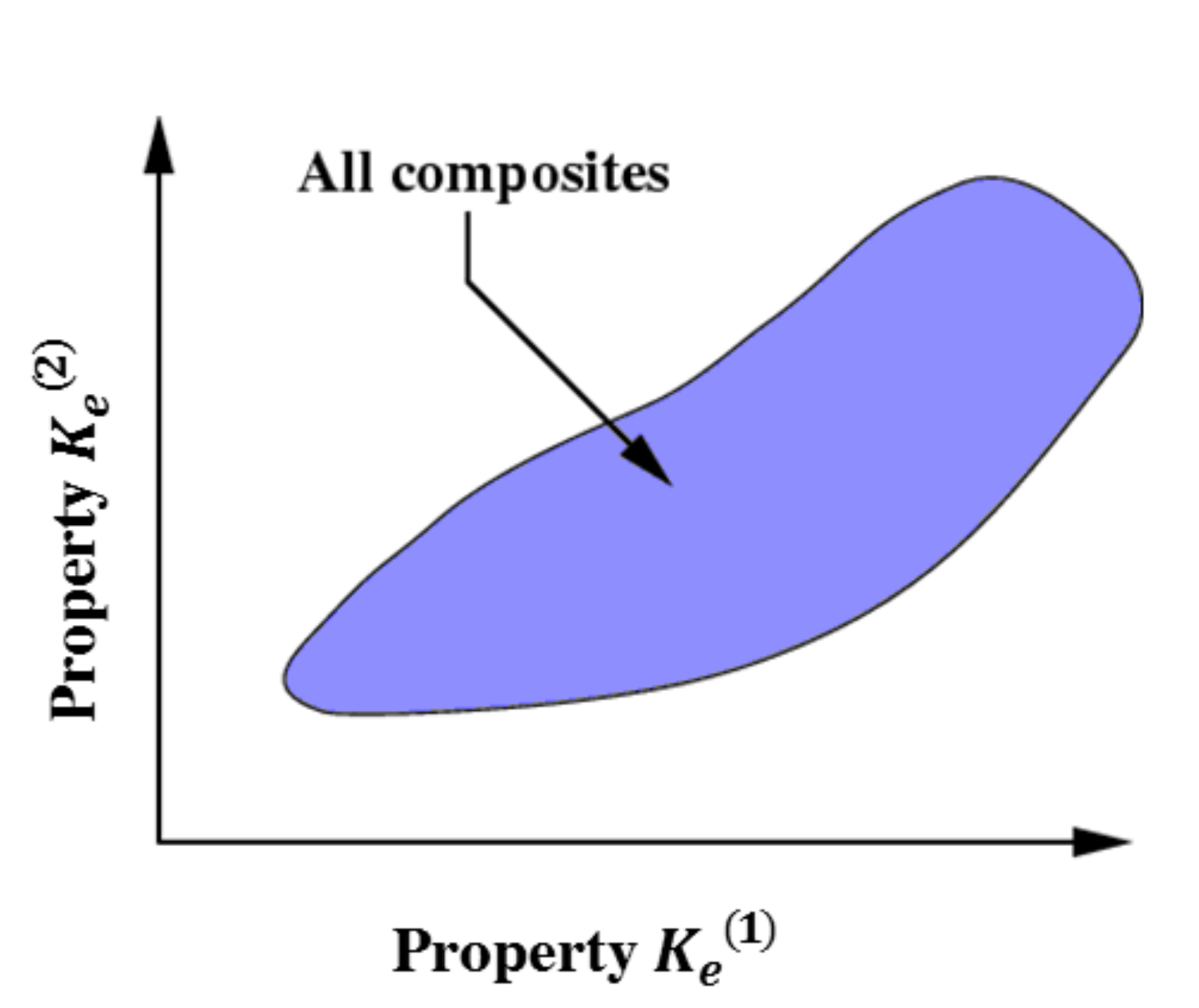}}\vspace{-0.15in}
\caption{ Schematic illustrating the allowable region in which all composites
with specified phase properties must lie for the case of two different effective properties, $K_e^{(1)}$ and $K_e^{(2)}$,
as adapted from Ref. \citenum{To02d}. Importantly, this allowable
region depends on the type of microstructural information that is specified.}
\label{multi}
\end{figure}

By contrast, composite materials are ideally suited to achieve
multifunctionality, since the best features of different materials
can be combined to form a new material that has a broad spectrum
of desired properties. \cite{Ch79,Ne93,To02a,Mi02,Sa03}
These materials may simultaneously perform as ultralight
load-bearing structures, enable thermal and/or electrical
management, ameliorate crash or blast damage, and
have desirable optical and acoustic characteristics. A general goal is the  design of
composite materials with $N$ different effective properties or responses,
which we denote by $K_e^{(1)},K_e^{(2)},\ldots, K_e^{(N)}$,
given the individual properties of the phases.
In principle, one desires to know the region (set)
in the multidimensional space of effective properties  in which
all composites must lie (see Fig. \ref{multi} for a two-dimensional (2D)
illustration). The size and shape of this region depends on the prescribed phase properties
as well as how much  microstructural information is specified, For example, the set of composites
with unspecified volume fractions  is clearly larger than
the set in which the the volume fractions are specified.

The determination of the allowable region is generally a highly
complex problem. Cross-property bounds \cite{Be78a,Av88,To90e,To91f,Av91b,Gi95b,Gi96b,Gi97c,To02a}
can aid to identify the boundary of the allowable region and numerical
topology optimization methods \cite{BeKik,Si97,To02d,Si03,To10b} can then be used to find specific microstructures
that lie on the boundary, which are extremal solutions. These methods often
bias the solutions to be periodic structures with high crystallographic symmetries.
As we will see below, it can be very beneficial to constrain the optimal solution set
to microstructures possessing ``correlated disorder", \cite{Yu21} which can have advantages
over periodic media,  especially an exotic type of disorder
within the so-called hyperuniformity class.\cite{To03a,To18a}
The purpose  of the present article is to review the emerging field
of disordered hyperuniform composites and their novel multifunctional characteristics.

\begin{figure*}
\centerline{\subfigure[]{\includegraphics[  width=1.8in,keepaspectratio,clip=]{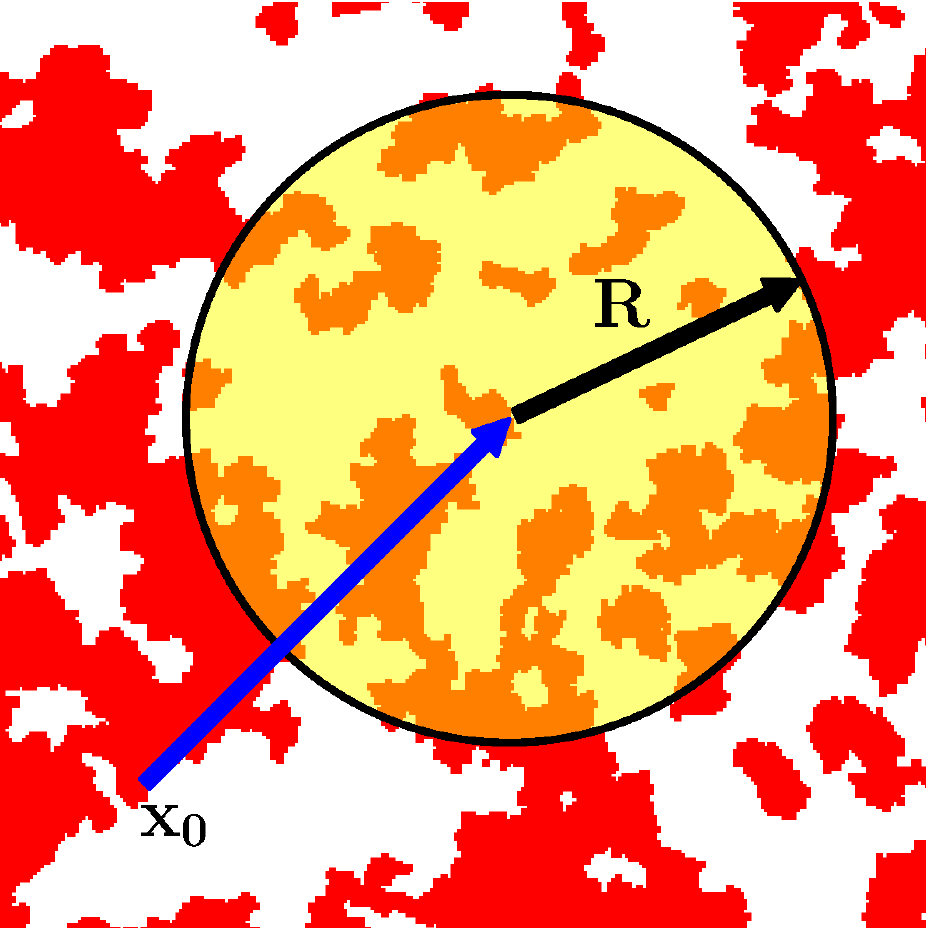}}\hspace{0.1in} \subfigure[]{\includegraphics[  width=1.8in,keepaspectratio,clip=]{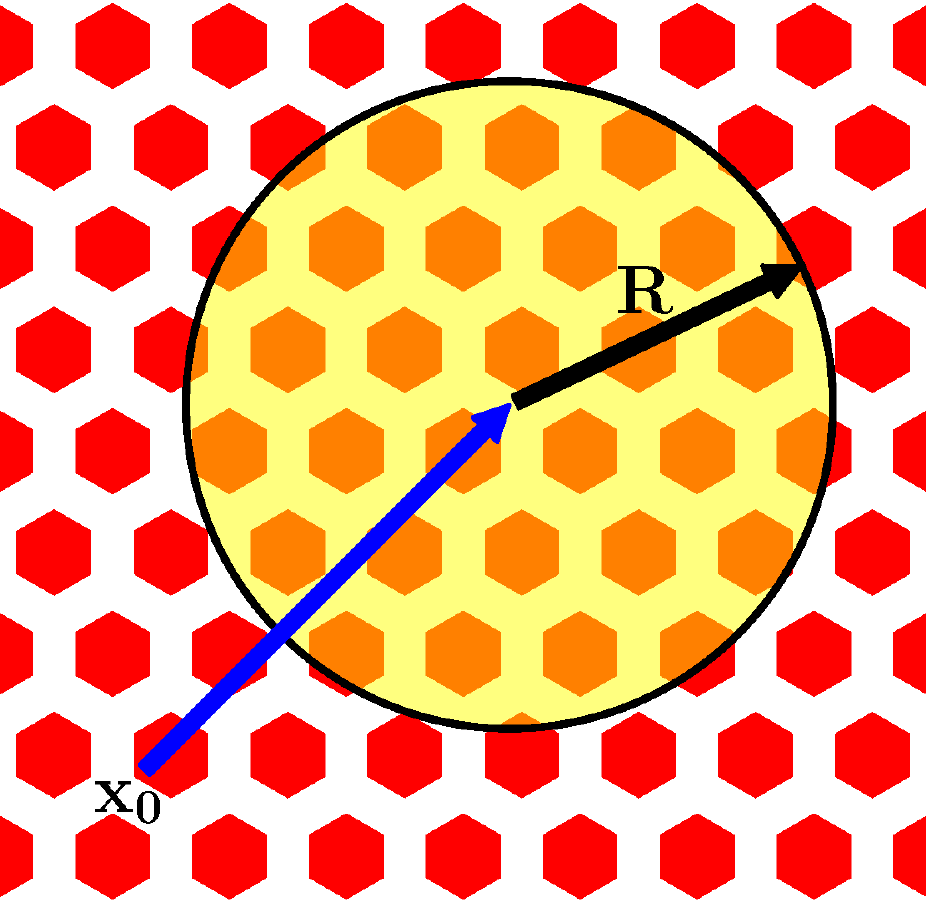}}
\hspace{0.1in}\subfigure[]{\includegraphics[  width=1.8in,keepaspectratio,clip=]{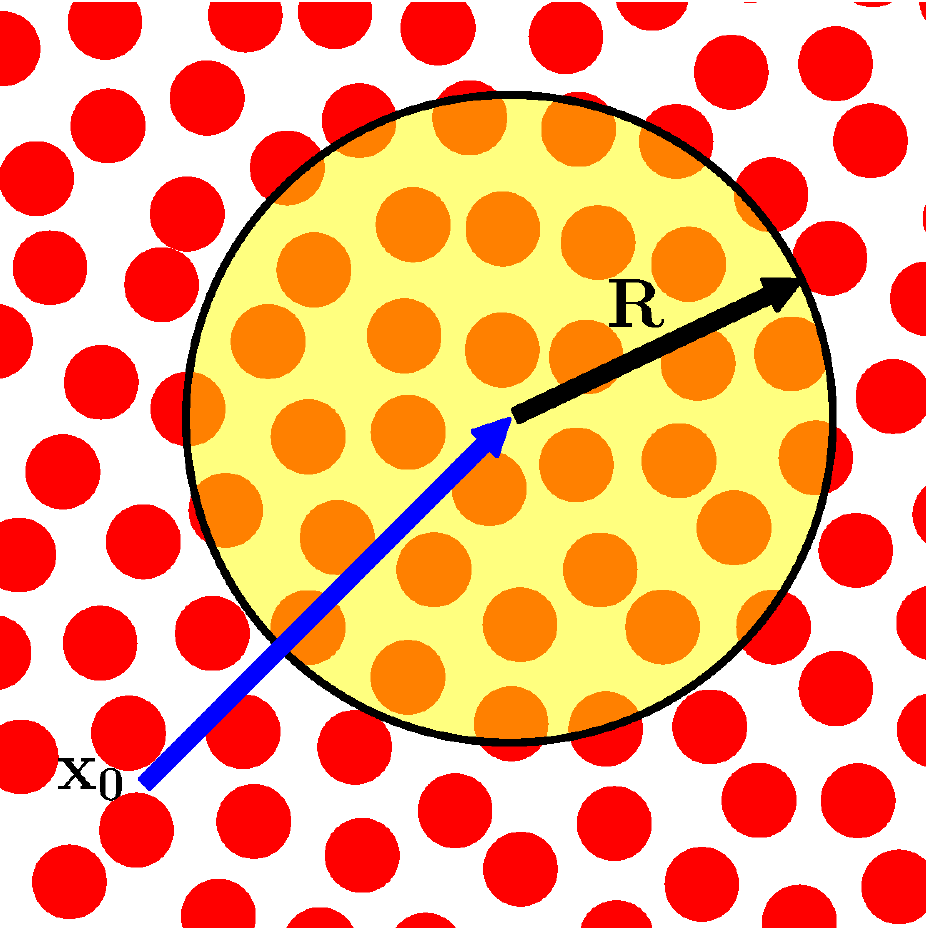}}}
\caption{Schematics indicating a circular observation window of radius $R$ in two dimensions and its centroid ${\bf x}_0$ for a 
``typical" disordered
nonhyperuniform (a), periodic (b), and disordered hyperuniform (c) media. In each of these examples, the
phase volume fraction within the window will fluctuate as the window position varies. Whereas the local variance $\sigma^2_{_V}(R)$
for the nonhyperuniform medium decays like $1/R^2$ for large $R$, it decays like $1/R^3$ in both the periodic and disordered hyperuniform
examples. In space dimension $d$ and for large $R$, $\sigma^2_{_V}(R)$ scales like $1/R^d$ and $1/R^{d+1}$  for nonhyperuniform
and hyperuniform media, respectively.}
\label{windows}
\end{figure*}

The hyperuniformity concept was introduced and studied nearly two decades ago
in the context of many-particle systems. \cite{To03a} Hyperuniform
systems are characterized by an anomalous suppression of large-scale density
fluctuations compared to ``garden-variety" disordered systems. The hyperuniformity concept generalizes the 
traditional notion of long-range order in many-particle systems
to not only include all perfect crystals and perfect quasicrystals, 
but also exotic amorphous states of matter. Disordered hyperuniform materials can have advantages over crystalline
ones, such as unique or nearly optimal, direction-independent
physical properties and robustness against defects. \cite{Fl09b,Ji14,De15,Le16,Ma16,Zh16b,Gk17,Fr17,Ch18a,Zh19,Go19,Sh20,Ki20a,Yu21}

The hyperuniformity concept was generalized to two-phase heterogeneous media in $d$-dimensional Euclidean space 
$\mathbb{R}^d$, \cite{Za09,To18a} which include composites, cellular solids and porous media.
A two-phase medium in $\mathbb{R}^d$ is hyperuniform if its local volume-fraction variance $\sigma^2_{_V}(R)$
associated with a spherical observation window of radius $R$ decays 
in the large-$R$ limit faster than the inverse of the window volume, i.e., $1/R^d$;
(see Sec. \ref{structure} for mathematical details).  This behavior is to be
contrasted with those of ``typical" disordered two-phase media
for which the variance decays like the inverse of the window volume (see Fig. \ref{windows}).

In this article, we review progress that has been made to generate and characterize 
multifunctional disordered two-phase composites and porous media. 
In Sec. \ref{structure}, we collect basic definitions and background on hyperuniform
and nonhyperuniform two-phase media. In Secs. \ref{forward}
and \ref{inverse}, we review developments in generating disordered hyperuniform two-phase media
using forward and inverse approaches, respectively. In Sec. \ref{order-metrics},
we describe order metrics that enable a rank ordering of disordered hyperuniform two-phase media.
 In Sec. \ref{novel-multi}, we review the current knowledge
about the extraordinary multifunctional characteristics of disordered hyperuniform composites and porous media.
Finally, In Sec. \ref{conclusions}, we make concluding remarks and discuss the outlook for the field.

\section{Structural Characterization of Disordered Hyperuniform Composites}
\label{structure}

\begin{figure}[H]
\subfigure[]{
\includegraphics[  width=2.75in, keepaspectratio,clip=]{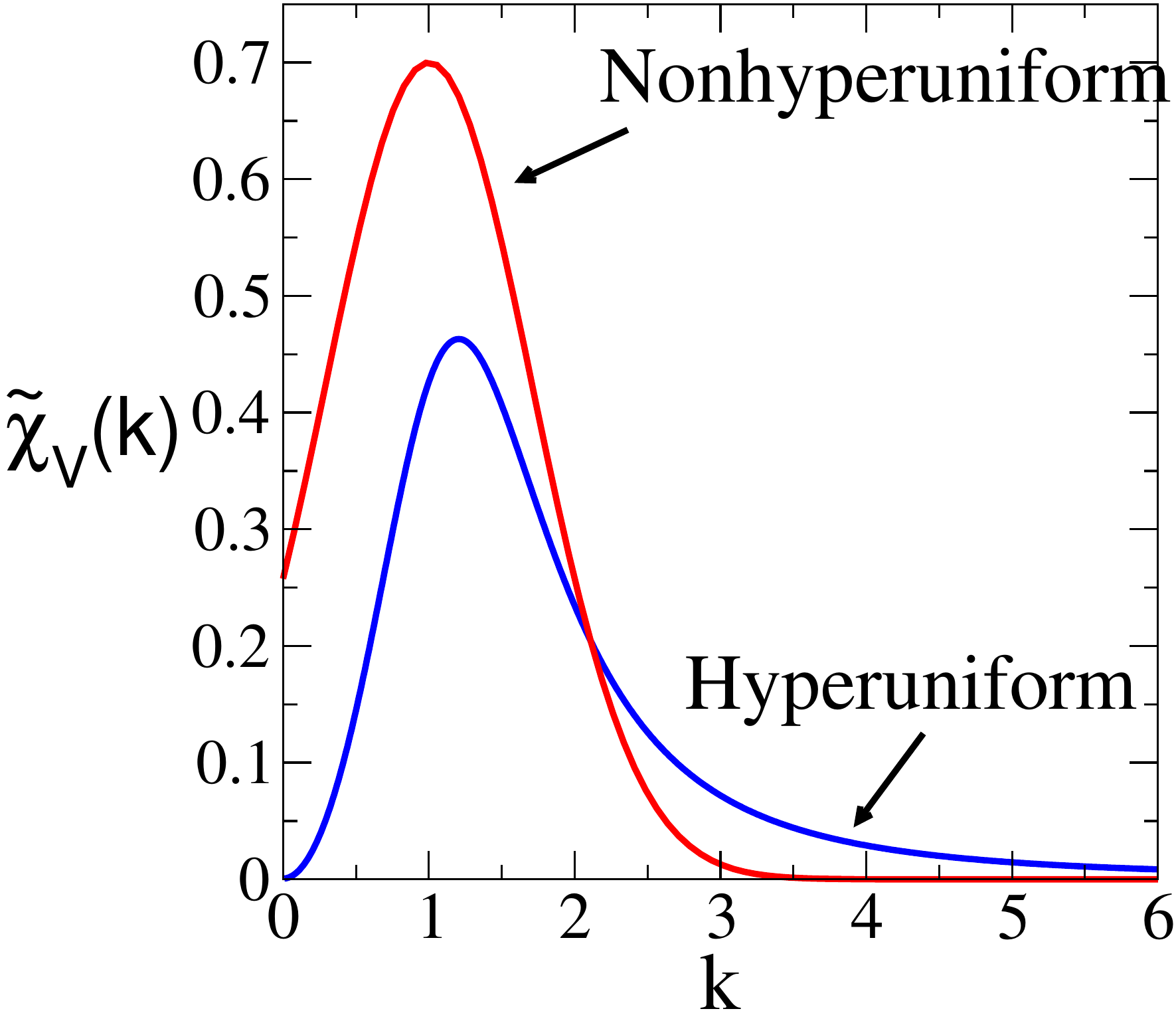}}\\
\subfigure[]{\includegraphics[  width=2.75in, keepaspectratio,clip=]{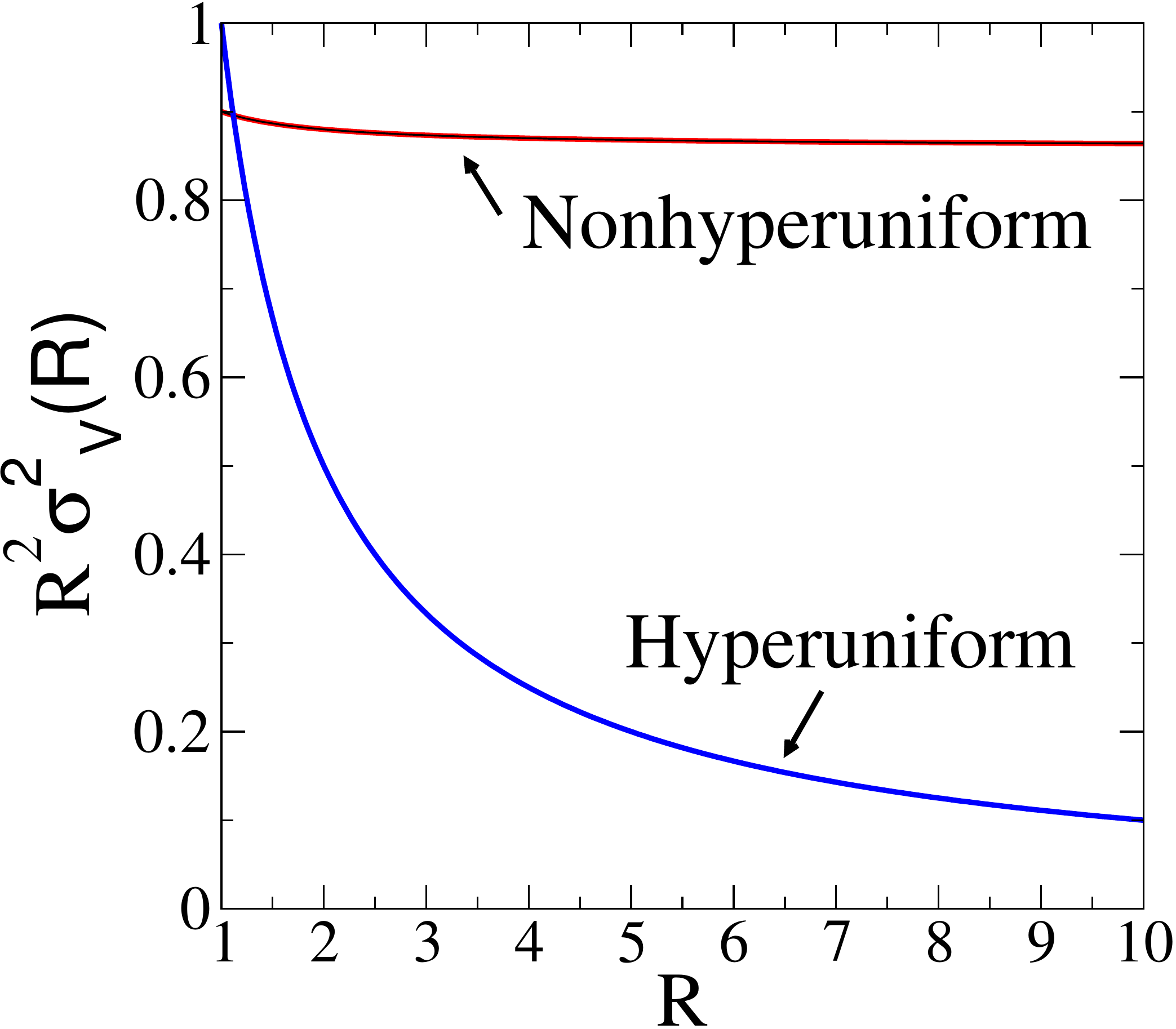}}
\caption{(a) Spectral densities versus wavenumber $k$ for 2D nonhyperuniform and hyperuniform media (b) Corresponding local variances (multiplied by $R^2$) versus window radius $R$.}
\label{examples}
\end{figure}

For two-phase heterogeneous media in $d$-dimensional Euclidean space $\mathbb{R}^d$, hyperuniformity 
can be  defined by the following infinite-wavelength  condition on the {\it spectral density} ${\tilde \chi}_{_V}({\bf k})$, \cite{Za09, To18a} i.e., 
\begin{equation}
\label{eq:HU_condition}
\lim_{|{\bf k}|\to 0 }{\tilde \chi}_{_V}({\bf k}) = 0,
\end{equation}
where $\bf k$ is the wavevector.
The spectral density ${\tilde \chi}_{_V}({\bf k})$ is the Fourier transform of the autocovariance function 
$\chi_{_V}({\bf r})\equiv S_2^{(i)}({\bf r})-{\phi_i}^2$, where $\phi_i$ is the volume fraction of phase $i$, and $S_2^{(i)}({\bf r})$ gives the probability of finding two points separated by $\bf{r}$ in phase $i$ at the same time. \cite{To02a}.
This two-point descriptor in Fourier space can be easily obtained for general microstructures either theoretically, computationally, or via scattering experiments. \cite{De57} The distinctions between the spectral densities for examples
of 2D hyperuniform and nonhyperuniform media can be vividly seen in the top panel of Fig. \ref{examples}.

Hyperuniformity of two-phase media can be also defined in terms of the {\it local volume-fraction variance} $\sigma^2_{_V}(R)$ associated with a spherical window of radius $R$. 
Specifically, a hyperuniform two-phase system is one in which $\sigma^2_{_V}(R)$ decays faster than $R^{-d}$ in the large-$R$ regime, \cite{Za09, To18a} i.e.,  
\begin{equation}
\label{eq:HU-condition2}
\lim_{R\to\infty}R^d \sigma^2_{_V}(R) = 0.
\end{equation}
In addition to having a direct-space representation, \cite{Lu90b} the local variance $\sigma^2_{_V}(R)$ has the following Fourier representation in terms
of the spectral density  ${\tilde \chi}_{_V}({\bf k})$ \cite{Za09,To18a}: 
\begin{equation}
\sigma^2_{_V}(R)  = \frac{1}{v_1(R)(2\pi)^d} \int_{\mathbb{R}^d} 
 {\tilde \chi}_{_V}({\bf k}) {\tilde \alpha}_2(k;R) d {\bf k}
, \label{eq:vv-Fourier}
\end{equation}
where $v_1(R)=\pi^{d/2}R^d/\Gamma(d/2+1)$ is the volume of a $d$-dimensional sphere of radius $R$, 
$\Gamma(x)$ is the gamma function, 
\begin{equation} \label{eq:alpha-tilde}
{\tilde \alpha}_2(k;R) \equiv 2^d \pi^{d/2} \Gamma(d/2+1)\frac{J_{d/2}(kR)^2}{k^d},
\end{equation} 
is the Fourier transform of the scaled intersection volume of two spheres of radius $R$ whose centers are separated by 
a distance $r$, \cite{To03a}
and $k \equiv |\bf k|$ is the wavenumber. The bottom panel of Fig. \ref{examples} depicts the local variances
corresponding to the spectral densities shown in the top panel.

As in the case of hyperuniform point configurations,  \cite{To03a, Za09}  there are three different scaling regimes (classes) that describe the associated large-$R$ behaviors of the volume-fraction variance when the spectral density goes to zero with the
following  power-law scaling: \cite{Za09,To16b,To18a}
\begin{equation}
{\tilde \chi}_{_V}({\bf k})\sim |{\bf k}|^\alpha \quad (k \to 0),
\label{scaling}
\end{equation}
namely,
\begin{align}   
\sigma^2_{_V}(R) \sim 
\begin{cases}
R^{-(d+1)}, \quad\quad\quad \alpha >1 \qquad &\text{(Class I)}\\
R^{-(d+1)} \ln R, \quad \alpha = 1 \qquad &\text{(Class II)},\\
R^{-(d+\alpha)}, \quad 0 < \alpha < 1\qquad  &\text{(Class III)}
\end{cases}
\label{eq:classes}
\end{align}
where the exponent $\alpha$ is a positive constant.
Classes I and III are the strongest and weakest forms of hyperuniformity, respectively.
Stealthy hyperuniform  media are also of class I and are defined to be those that possess
zero-scattering intensity for a set of wavevectors around the origin, \cite{To16b} i.e.,
\begin{equation}
{\tilde \chi}_{_V}({\bf k})=0 \qquad \mbox{for}\; 0 \le |{\bf k}| \le K.
\label{stealth}
\end{equation}
Examples of such media are periodic packings of spheres
as well as unusual disordered sphere packings derived from stealthy point patterns. \cite{To16b,Zh16b}

By contrast, for any nonhyperuniform two-phase system, it is straightforward to show,
using a similar analysis as for point configurations, \cite{To21c} that
the local variance has the following large-$R$ scaling behaviors:
\begin{align}  
\sigma^2_{_V}(R) \sim 
\begin{cases}
R^{-d}, & \alpha =0 \quad \text{(typical nonhyperuniform)}\\
R^{-(d+\alpha)}, & -d <\alpha < 0 \quad \text{(antihyperuniform)}.\\
\end{cases}
\label{sigma-nonhyper}
\end{align}
For a  ``typical" nonhyperuniform system, ${\tilde \chi}_{_V}(0)$ is bounded.  \cite{To18a} In {\it antihyperuniform} systems,
${\tilde \chi}_{_V}(0)$ is unbounded, i.e.,
\begin{equation}
\lim_{|{\bf k}| \to 0} {\tilde \chi}_{_V}({\bf k})=+\infty,
\label{antihyper}
\end{equation}
and hence  are diametrically opposite to hyperuniform systems.
Antihyperuniform systems include  systems at thermal critical points (e.g., liquid-vapor and magnetic critical points), \cite{St87b,Bi92} fractals, \cite{Ma82} disordered non-fractals, \cite{To21b}
and certain substitution tilings. \cite{Og19}

\section{Forward Approaches to Generating Disordered Hyperuniform Two-Phase Media}
\label{forward}

\begin{figure*}[bt]
\centering
\subfigure[\ Disks]{
\includegraphics[width=5.cm, height=5.cm]{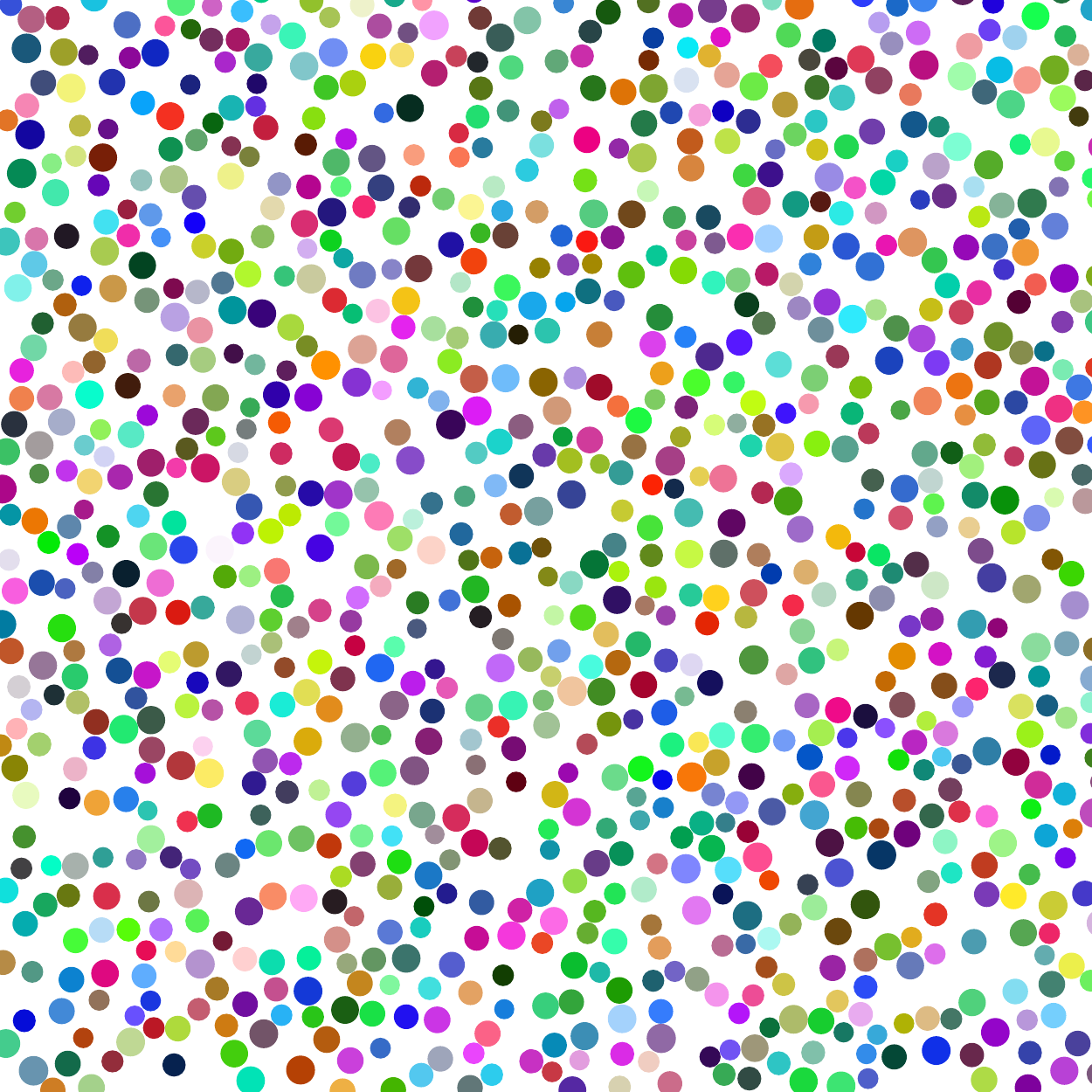}
}
\subfigure[\ Needles]{
\includegraphics[width=5.cm, height=5.cm]{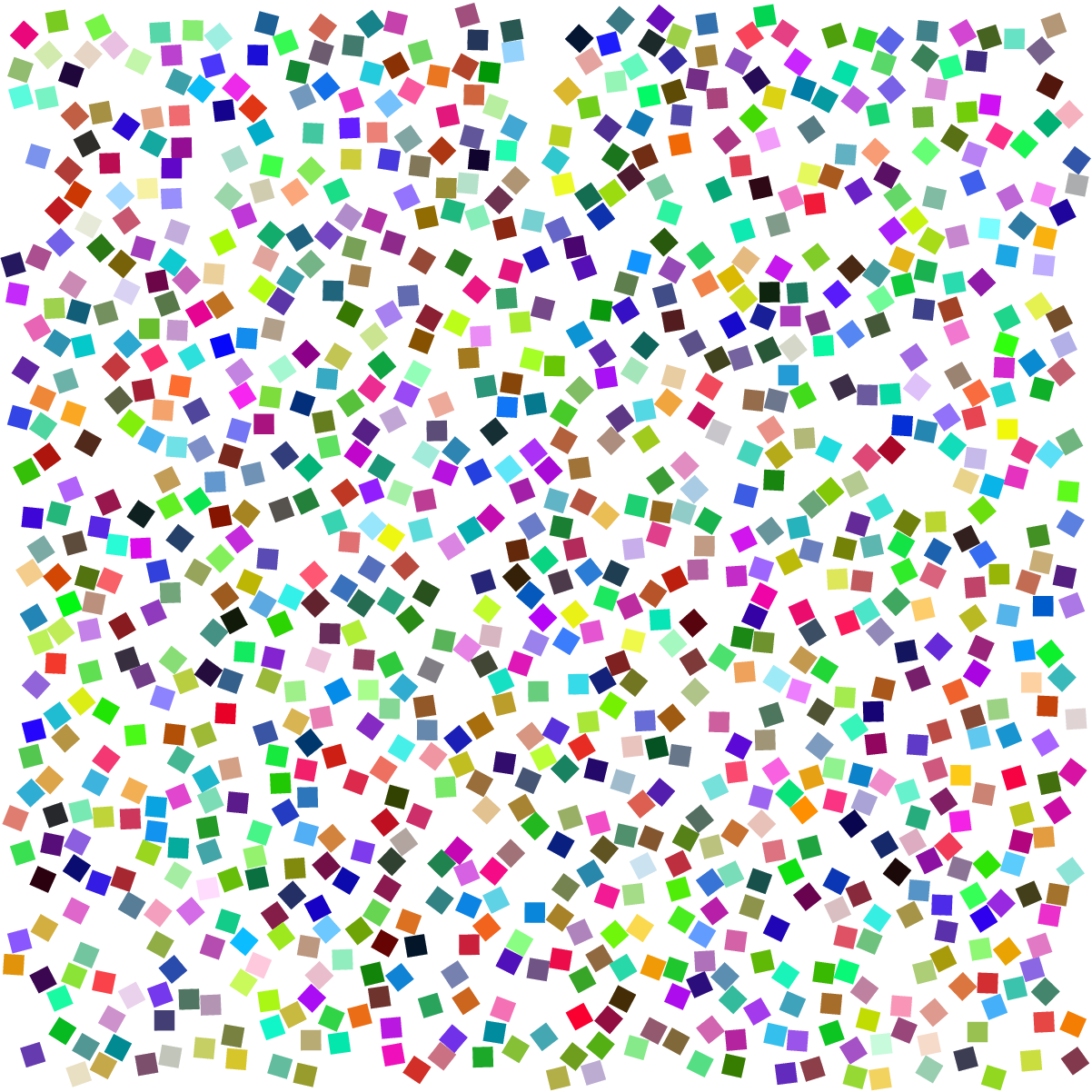}
}
\subfigure[\ Squares]{
\includegraphics[width=5.cm, height=5.cm]{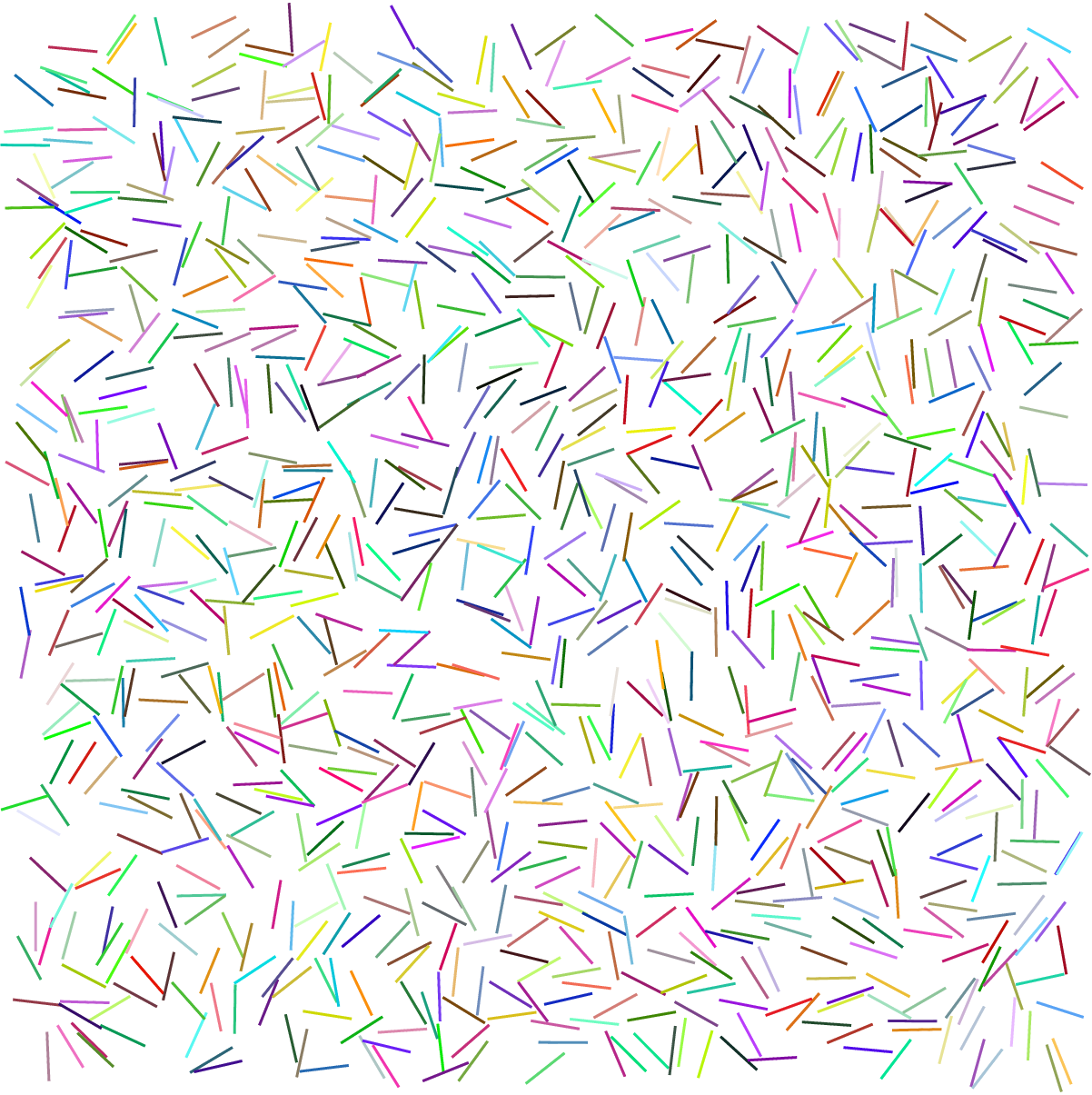}
}
\caption{ Representative images of 2D hyperuniform absorbing-state particle configurations, as
adapted from Ref. \citenum{Ma19}. (a) Disks with continuous size distribution. (b) Identical needles. (c) Identical squares. }
\label{nonspherical}
\end{figure*}

Here we describe ``forward" (direct) approaches that have yielded disordered hyperuniform particulate media.
Jammed as well as unjammed states are briefly reviewed.

Torquato and Stillinger \cite{To03a} suggested
that certain defect-free strictly jammed (i.e., mechanically stable) packings of identical
spheres are hyperuniform. Specifically, they conjectured that
any strictly jammed saturated infinite packing of identical
spheres is hyperuniform. A saturated packing of hard spheres
is one in which there is no space available to add another
sphere. This conjecture was confirmed by Donev et al. \cite{Do05d}
via a numerically generated maximally random jammed (MRJ) packing \cite{To00b,To18b} of 
$10^6$ hard spheres in three dimensions. Subsequently, the hyperuniformity
of other MRJ hard-particle packings, including nonspherical particle shapes,
was established across dimensions. \cite{Sk06,Ji10b,Za11a,Ji11c,Ho12b,Ch14a,Ti15,Kl16,At16a,At16b,Cin18,Ma22a}
Jammed athermal soft-sphere models of granular media, \cite{Si09,Be11} jammed thermal colloidal packings, \cite{Ku11,Dr15}
and jammed bidisperse emulsions \cite{Ri17} were also shown to be effectively hyperuniform.
The singular transport and electromagnetic properties of MRJ packings of spheres \cite{Kl18} and superballs \cite{Ma22a}
have also been investigated.

Periodically driven colloidal suspensions were observed to have a phase transition in terms of the reversibility of the dynamics one decade ago \cite{Pi05}. Random organization models capture the salient physics of how driven systems can self-organize. \cite{Chaik08}
A subsequent study of random organization models of monodisperse (i.e., identical) spherical particles have shown that
a hyperuniform state is achievable when a granular system goes through an absorbing phase transition to a critical
state. \cite{He15} 
Many variants of such models and systems have been studied numerically. \cite{He17a, Tj15,Di15,Go17,Wa18}
To what extent is hyperuniformity is preserved when the model is generalized to particles
with a size distribution and/or nonspherical shapes? This question was probed in a recent study
\cite{Ma19} by examining  disks with a size
distribution, needle-like shapes and squares in two dimensions and it was
demonstrated that their critical states are hyperuniform as two-phase media (see Fig. \ref{nonspherical}).
These results suggest that general particle systems
subject to random organization can be a robust way to fabricate a wide class of hyperuniform states of matter
by tuning the structures via different particle-size and -shape distributions. This tunability capacity in turn potentially enables the
creation of multifunctional hyperuniform materials with desirable optical, transport, and mechanical properties.

\begin{figure*}
\begin{center}
\subfigure[]{
\includegraphics[width=2in]{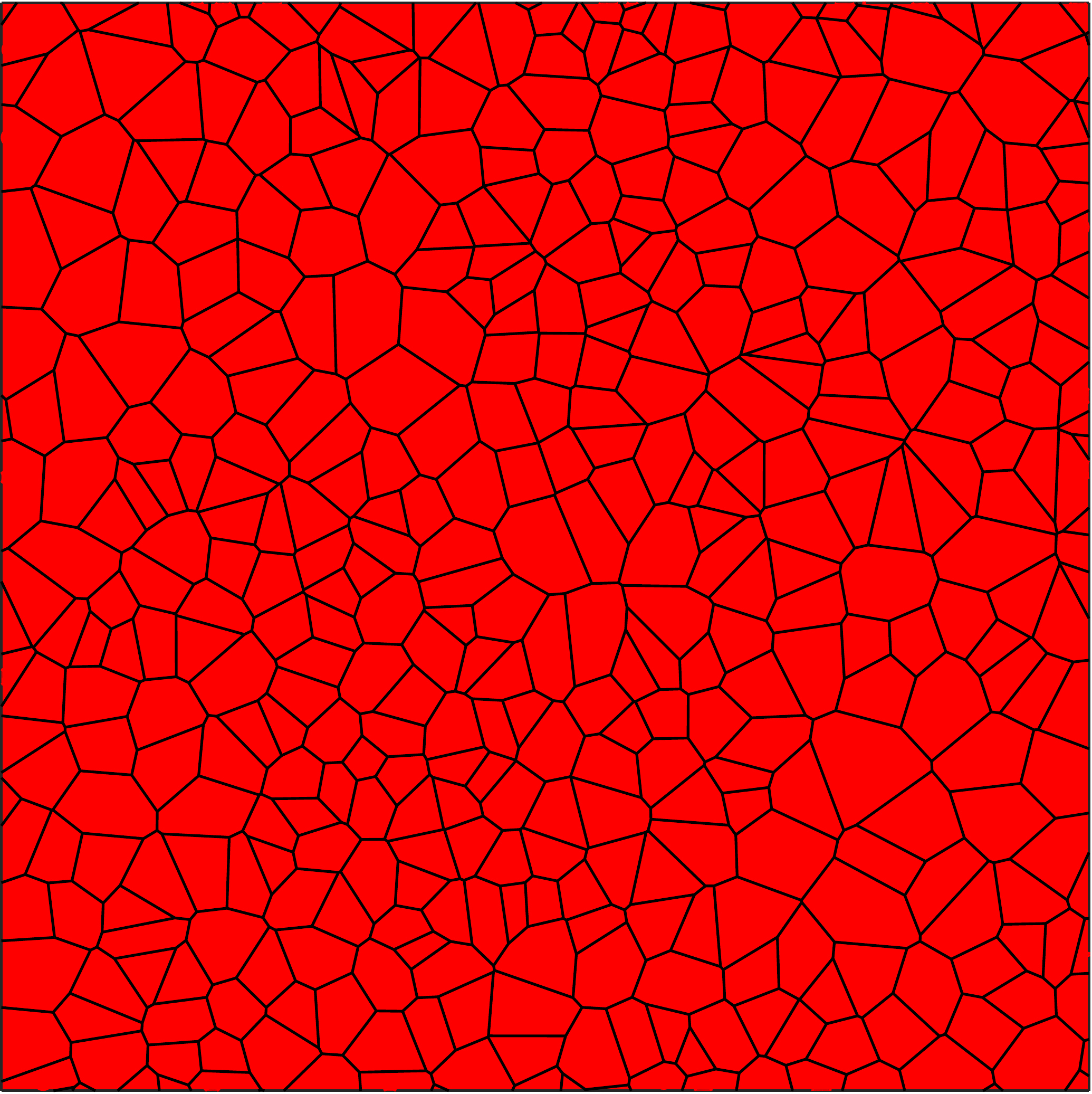}}
       \hspace{10pt}
   \subfigure[]{     
\includegraphics[width=2in]{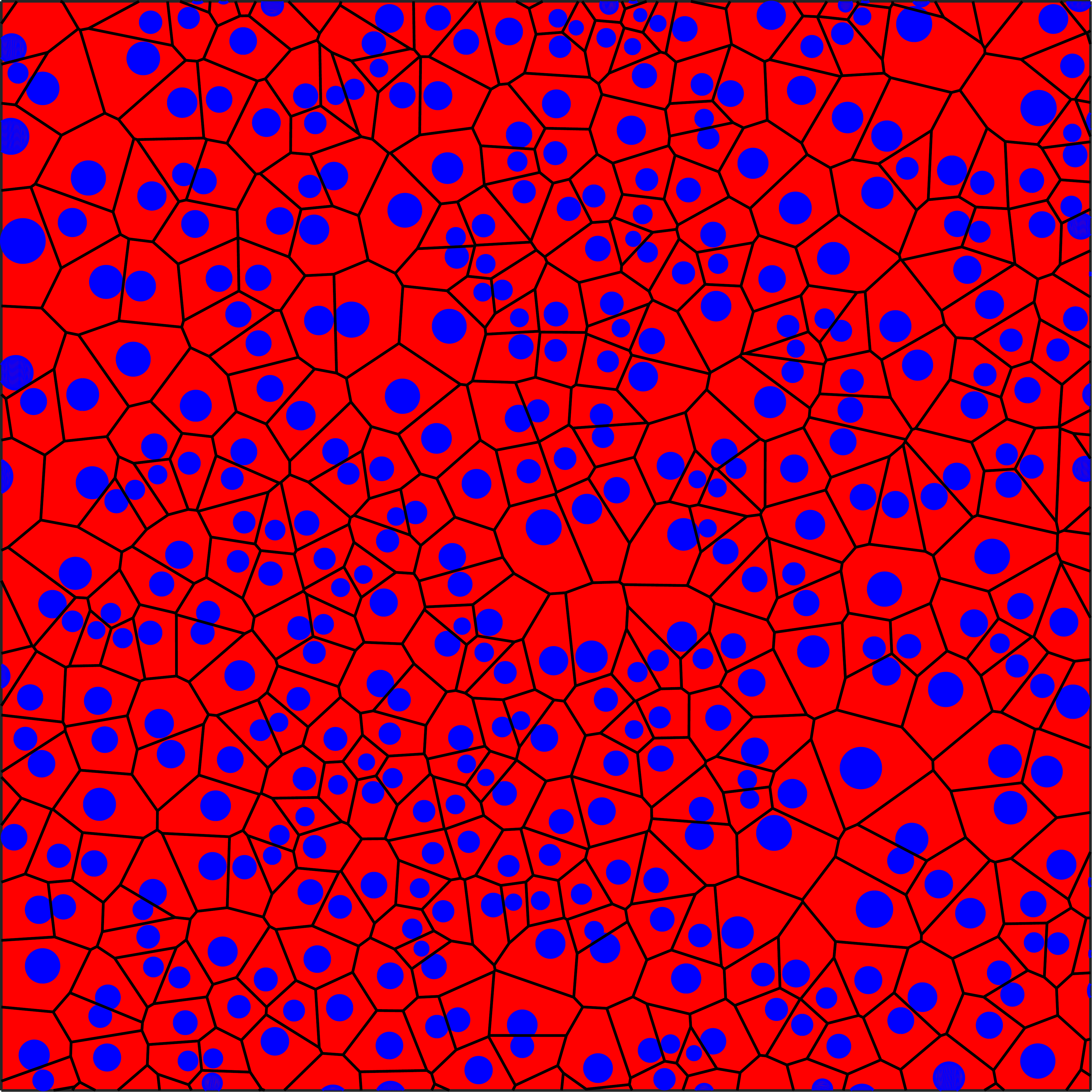}}\\
\subfigure[]{
\includegraphics[width=2in]{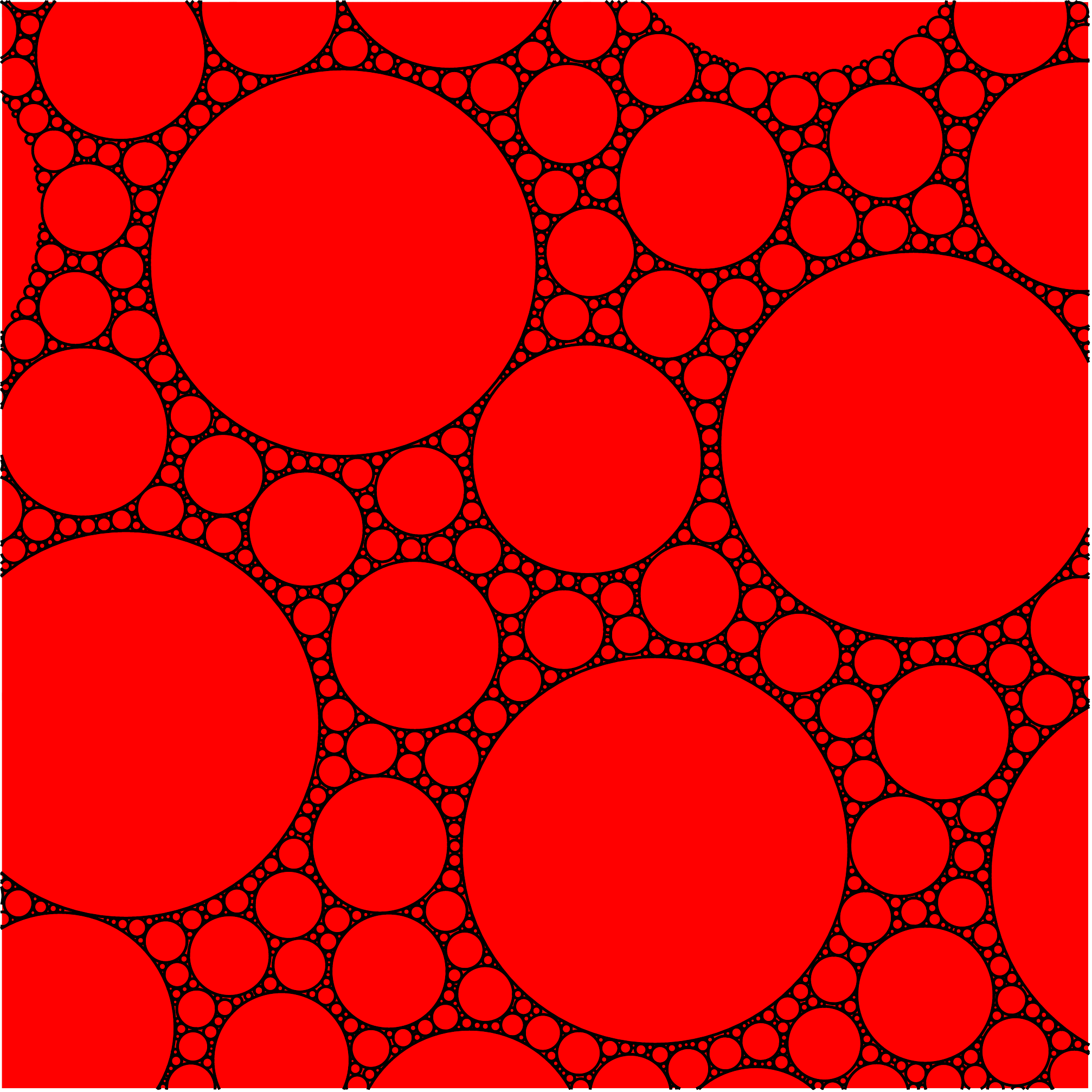}}
       \hspace{10pt}
\subfigure[]{
        \includegraphics[width=2in]{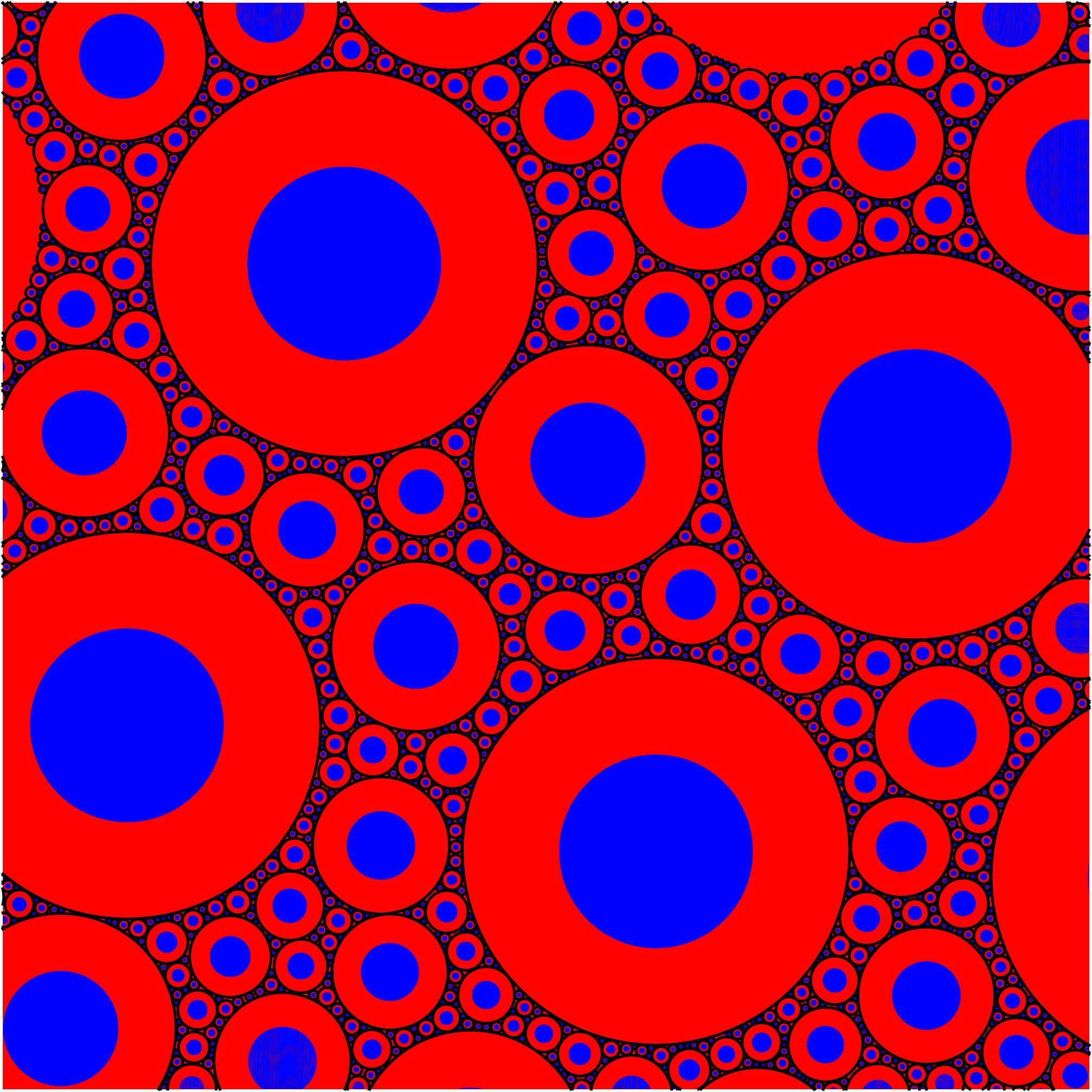}}
\end{center}
\caption{(a) A Voronoi tesselation of a nonhyperuniform disordered point configuration. (b)
The disordered hyperuniform packing of spheres with a size distribution that results by adding particles in the Voronoi tesselation
while ensuring that the local cell packing fraction
is equal to global packing fraction. (c) A tessellation of space by spheres. (d) The Hashin-Shtrikman composite sphere assemblage that results by adding particles in the sphere
tessellation while  ensuring that the local cell packing fraction
is equal to global packing fraction. These images are adapted from those in Ref. \citenum{Ki19b}.}
\label{tiling}
\end{figure*}

\begin{figure*}[bthp]
\begin{center}
\subfigure[]{
 \includegraphics[width = 0.35\textwidth,clip=]{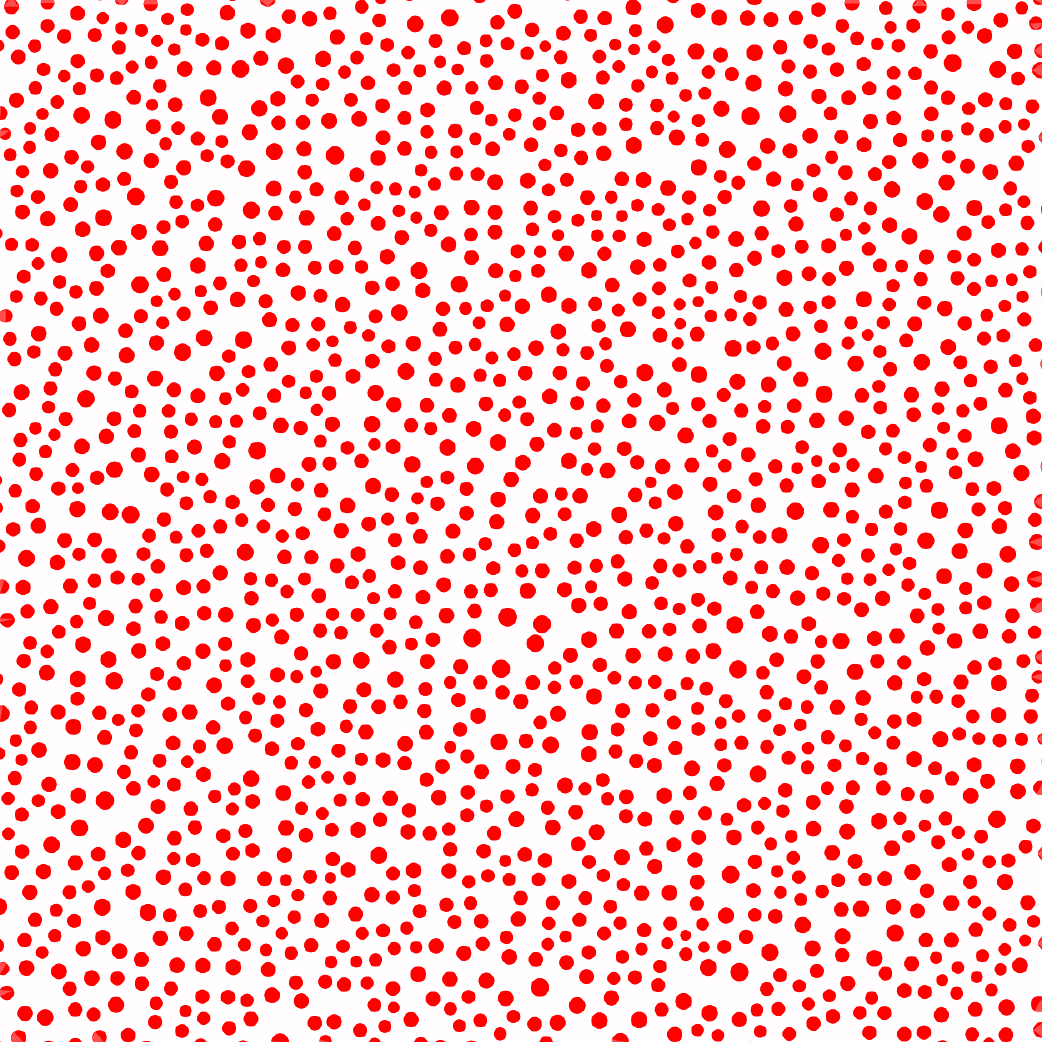}}\hspace{0.35in}
 \subfigure[]{
 \includegraphics[width = 0.35\textwidth,clip=]{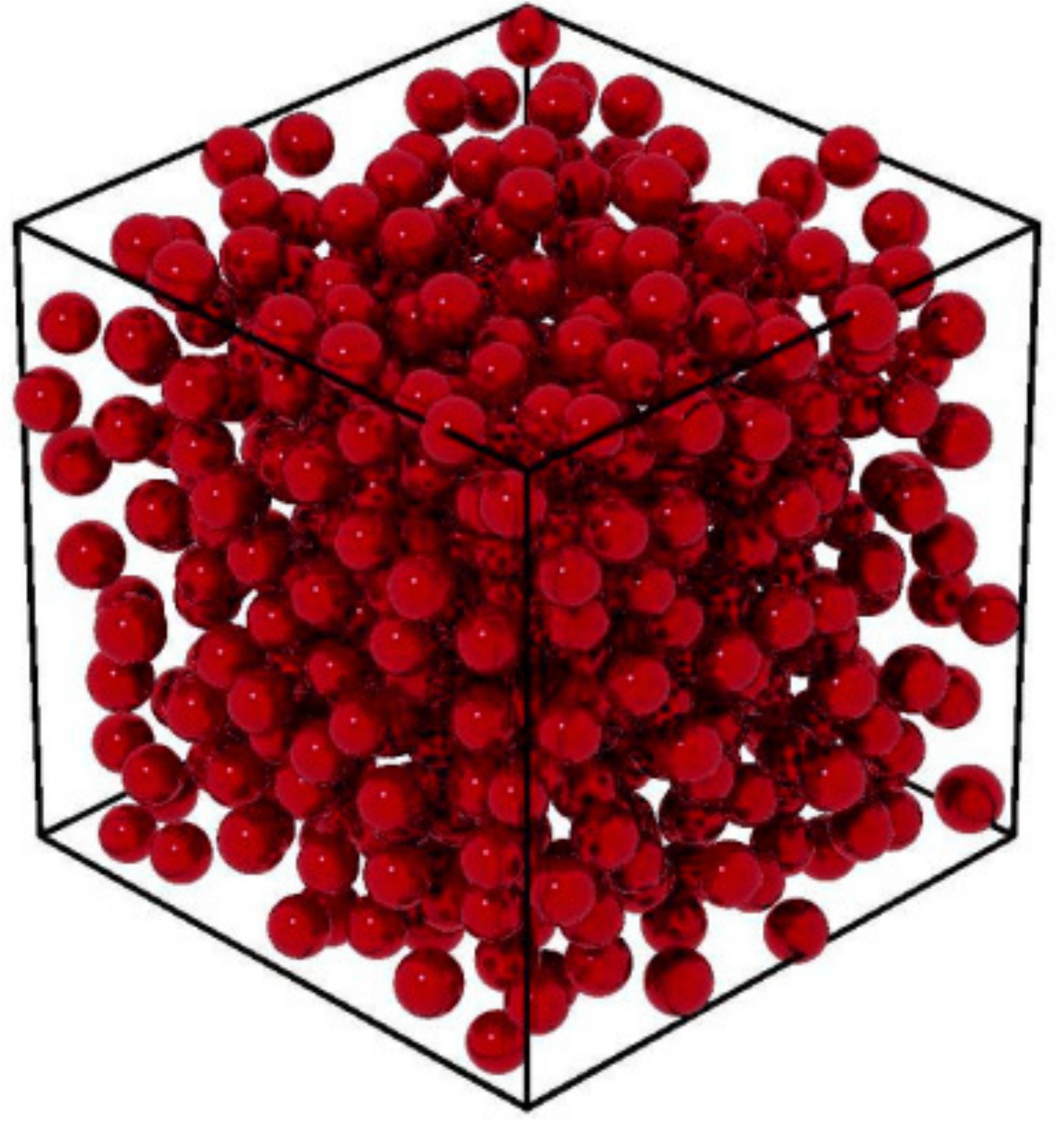}}
\end{center}
\caption{(a) A portion of a hyperuniform disk
packing that was converted from a 2D RSA packing with the
packing fraction $\phi_{init} = 0.41025$. (b) A portion of a 
hyperuniform sphere packing that was converted from a 3D saturated
RSA packing with the packing fraction $\phi_{init} = 0.288$. This figure is adapted from  Ref. \citenum{Ki19b}}
\label{RSA}
\end{figure*}

While there has been growing interest in disordered hyperuniform materials states, 
an obstacle has been an inability to produce large samples that are perfectly
hyperuniform due to practical limitations of conventional numerical and experimental methods. To
overcome these limitations, a general theoretical methodology has been developed to construct perfectly
hyperuniform packings in $d$-dimensional Euclidean space $\mathbb{R}^d$.\cite{Ki19a,Ki19b} Specifically, beginning with an initial
general tessellation of space by disjoint cells that meets a ``bounded-cell” condition, 
hard particles are placed inside each cell such that the local-cell particle packing fractions are identical to
the global packing fraction; see Fig. \ref{tiling}. It was proved that the constructed packings with a polydispersity in size
in $\mathbb{R}^d$ are perfectly hyperuniform of class I in the infinite-sample-size limit, even though 
the initial point configuration that underlies the Voronoi tessellation is nonhyperuniform.
Implementing this methodology in sphere tessellations of space (requiring spheres down to the infinitessimally  small), 
establishes the hyperuniformity of the classical
Hashin-Shtrikman multiscale coated-spheres structures, which are known to be two-phase media
microstructures that possess optimal effective transport and elastic properties. \cite{Ha62c,Ha63}
Figure  \ref{RSA} shows portions of 2D and 3D hyperuniform polydisperse packings 
that were converted from the corresponding Voronoi tessellations of  nonhyperuniform random sequential addition (RSA)
packings \cite{To06d}. These computationally-designed microstructures can be fabricated via either
photolithographic and 3D-printing techniques. \cite{Wo12,Va13,Sh15,Zhao18}

\section{Inverse Approaches to Generating Disordered Hyperuniform Two-Phase Media}
\label{inverse}

Here, we describe inverse optimization techniques that enable one to design microstructures
with targeted spectral densities. These procedures include the capacity
to tune the value of the power-law exponent $\alpha>0$, defined by relation (\ref{scaling}),
for nonstealthy hyperuniform media as well as to design stealthy hyperuniform media, defined by relation (\ref{stealth}).

The Yeong-Torquato stochastic optimization procedure \cite{Ye98a, Ye98b} 
is a popular algorithm that has been employed to construct or reconstruct digitized multi-phase media from 
a prescribed set of different correlation functions in physical (direct) space. \cite{Ji07,Ji09b,Pa15,Xu17,Ka18,Ca18,Li18,Sk21}
A fictitious "energy" is defined to be a sum of squared differences between
the target and simulated correlation function.  The Yeong-Torquato
procedure treats the construction or reconstruction task as an energy-minimization problem
that it solves via simulated annealing.
The Yeong-Torquato procedure was generalized to construct disordered hyperuniform materials with desirable effective macroscopic properties but from targeted structural information in Fourier (reciprocal) space, namely, the spectral density $\tilde{\chi}_{_V}({\bf k})$.\cite{Ch18a} Specifically, the fictitious ``energy'' $E$ of the system in $d$-dimensional Euclidean space $\mathbb{R}^d$
is defined as the following sum over wavevectors:
\begin{equation}
\label{energy} 
E = \sum_{\bf k} [\widetilde{\chi}_{_V}({\bf k})/l^d - \widetilde{\chi}_{_{V,0}}({\bf k})/l^d]^2,
\end{equation}
where the sum is over discrete wave vectors ${\bf k}$, $\widetilde{\chi}_{_{V,0}}({\bf k})$ and $\widetilde{\chi}_{_V}({\bf k})$ are the spectral densities of the target and (re)constructed microstructures, respectively, $d$ is the space dimension, and $l$ is the relevant characteristic length of the system used to scale the spectral densities such that they are dimensionless. As in the standard Yeong-Torquato
procedure, \cite{Ye98a,Ye98b} the simulated-annealing method is used to minimize the energy  (\ref{energy}).
It was demonstrated that one can design nonstealthy hyperuniform media and stealthy hyperuniform media
with this Fourier-based inverse technique \cite{Ch18a}. Such {\it in-silico} designed microstructures can be readily realized by 3D printing and lithographic technologies. \cite{Va13}

\begin{figure*}[bthp]
\centerline{
\includegraphics[  width=5in, keepaspectratio,clip=]{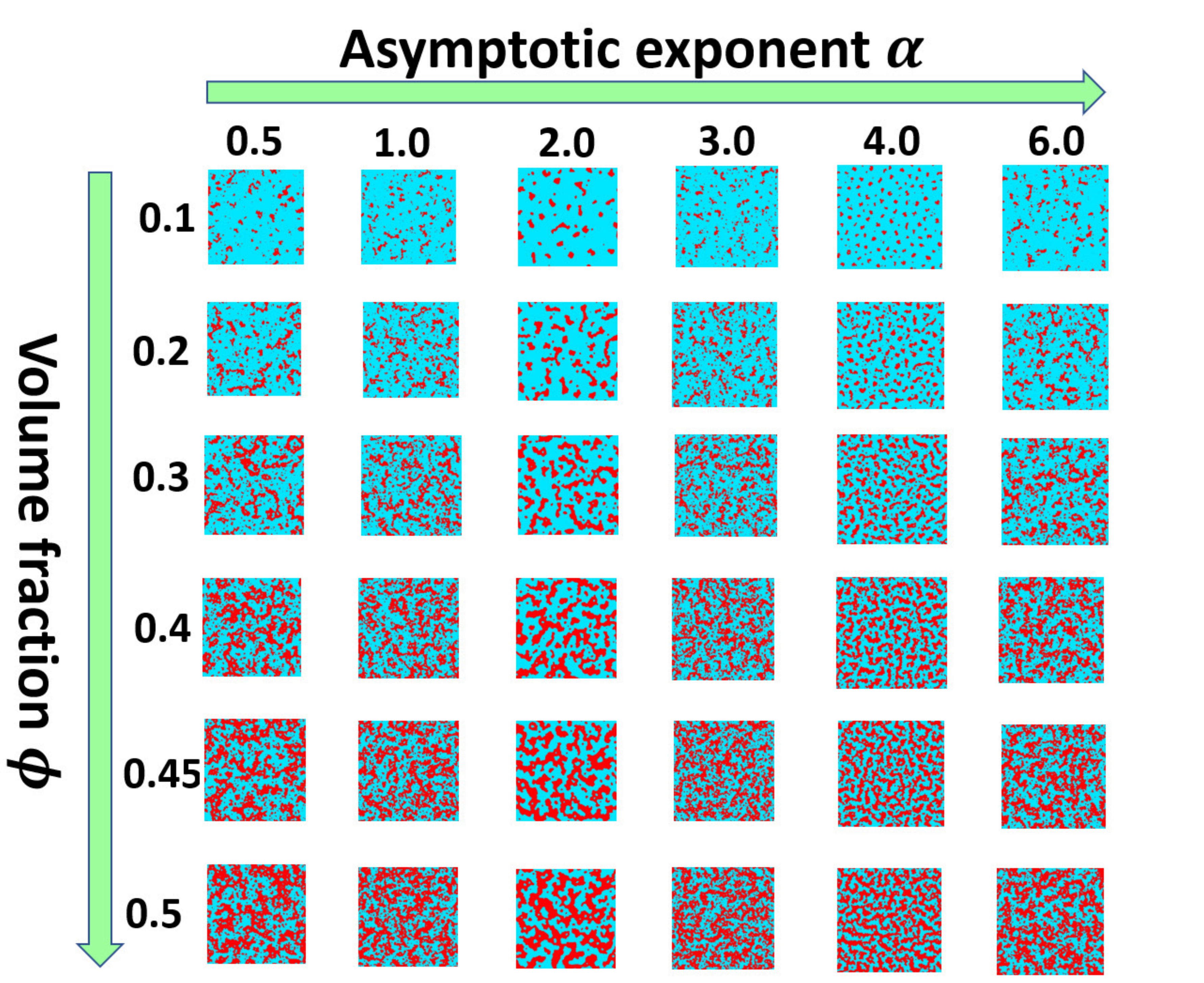} }
\caption{Realizations of disordered hyperuniform two-phase materials for different values of 
the volume fraction $\phi$ and the positive exponent  $\alpha$, defined by the spectral-density scaling law (\ref{scaling}).
This figure is adapted from Ref. \citenum{Ch18a}.} 
\label{designs}
\end{figure*}

Figure \ref{designs} shows designed realizations of digitized nonstealthy hyperuniform media with prescribed  values of the power-law exponent $\alpha>0$, defined 
by relation (\ref{scaling}), at different values of the phase volume fraction $\phi$. \cite{Ch18a} 
It is seen that these designed materials possess a variety of morphologies: as $\phi$ increases for fixed $\alpha$,
the microstructures transition from particulate media consisting of isolated ``particles" to
labyrinth-like microstructures. Moreover, as $\alpha$ increases for fixed $\phi$, short-range order
increases in these hyperuniform materials.

The ``collective-coordinate" optimization procedure represents a powerful reciprocal-space-based approach 
to generate disordered stealthy hyperuniform point configurations  \cite{Uc04b,Ba08,Zh15a}
as well as nonstealthy hyperuniform point configurations \cite{Uc06b,Zh16a} in $d$-dimensional Euclidean space 
$\mathbb{R}^d$. In the case of the former, their degree
of the stealthiness can be tuned by varying a dimensionless parameter $\chi$, which measures the
relative number of independently constrained degrees of
freedom for wavenumbers up to the cut-off value $K$. For the range    $0 < \chi <1/2$,
the stealthy states are disordered, and degree of short-range order increases
as $\chi$ approaches $1/2$. \cite{To15} Such stealthy point patterns
can be decorated by nonoverlapping spheres, enabling the generation
of {\it non-digitized} stealthy sphere packings, which have been used to model disordered
two-phase composites that are both stealthy and hyperuniform, \cite{Zh16b,Ki20a} as defined by relation
(\ref{stealth}). As we will see in Sec. \ref{novel-multi}, stealthy hyperuniform composites
are endowed with novel multifunctional characteristics.

\section{Order Metrics for Disordered Hyperuniform Two-Phase Media}
\label{order-metrics}

An outstanding open problem is the determination  of appropriate ``order metrics" to characterize the degree 
of large-scale order of both hyperuniform and nonhyperuniform media.
This task is a highly challenging  due to the infinite variety of possible two-phase microstructures (geometries and topologies). 
To begin such a program, the local variance $\sigma^2_{_V}(R)$ was recently studied for a certain subset
of class I hyperuniform media, including 2D  periodic cellular networks as well as 2D periodic and disordered/irregular packings, some of which maximize their effective transport and elastic properties. \cite{Ki21}
In particular, Kim and Torquato \cite{Ki21} evaluated the local variance $\sigma^2_{_V}(R)$ as a function of the window radius $R$.
They also computed  the hyperuniformity order metric $\overline{B}_V$, i.e.,  the implied coefficient multiplying $R^{-(d+1)}$ in (\ref{eq:classes}),
for all of these class I 2D models to rank  them according to their degree of order at a fixed volume fraction. The smaller is the value of  $\overline{B}_V$, the more
ordered is the microstructure with respect to large-scale volume-fraction fluctuations.
Among the cellular networks considered, the
honeycomb networks have the minimal values of the hyperuniformity order metrics $\overline{B}_V$ across all volume fractions.
Among all structures studied there,  triangular-lattice
packings of circular disks have the minimal values of the order metric for almost all volume fractions.

The extension of the aforementioned work to other 2D hyperuniform microstructures as well as to 3D hyperuniform media
are important avenues for future research. Such investigations of  a wider array of hyperuniform media
would be expected to  lead to improved order metrics and facilitate materials discovery.

\section{Novel Multifunctional Disordered Hyperuniform Composites and Porous Media}
\label{novel-multi}

\begin{figure*}[bthp]
\centerline{\subfigure[]{\includegraphics[  width=3.5in, keepaspectratio,clip=]{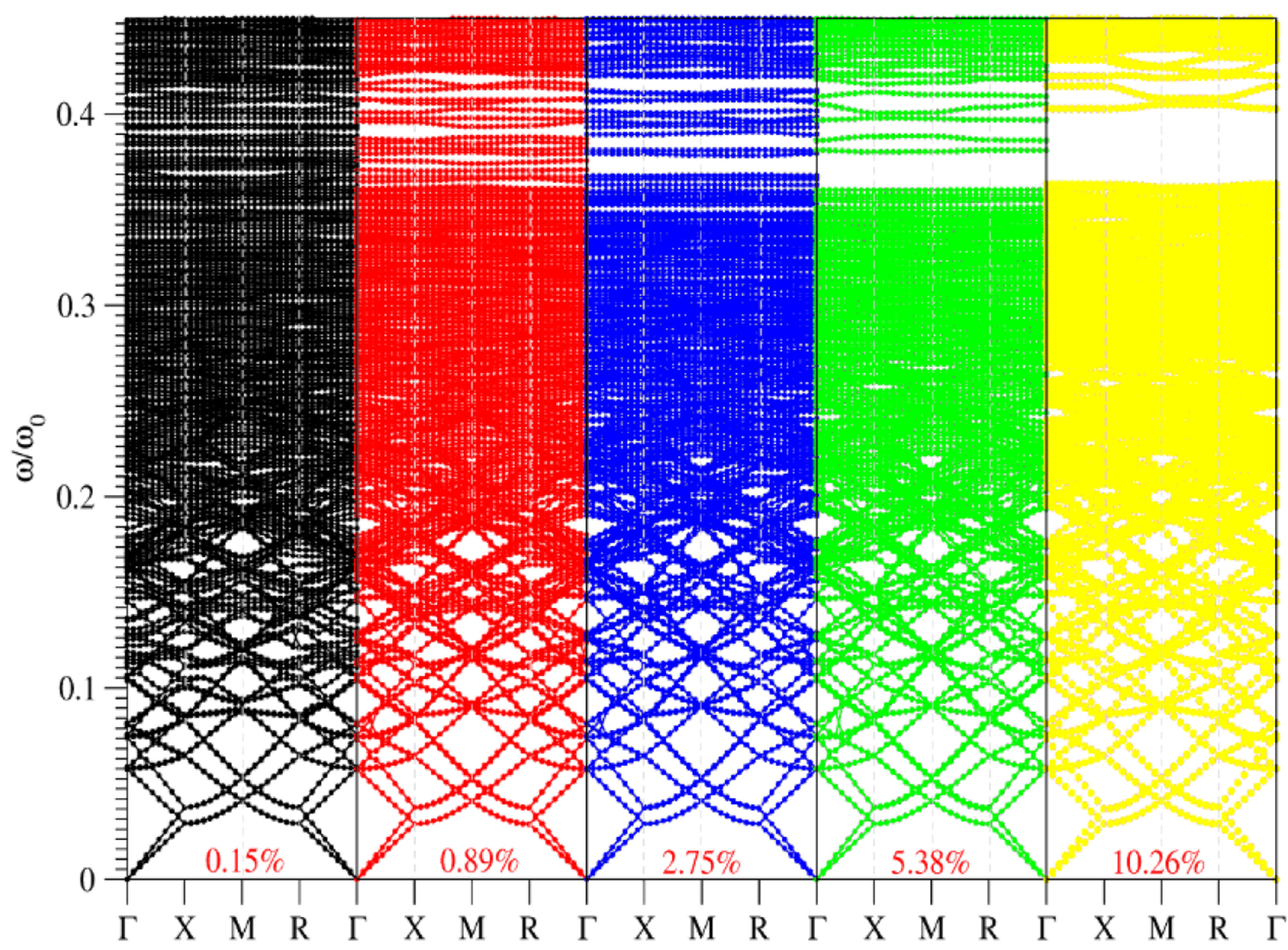}}\hspace{0.2in}\subfigure[]{\includegraphics[  width=2.3in, keepaspectratio,clip=]{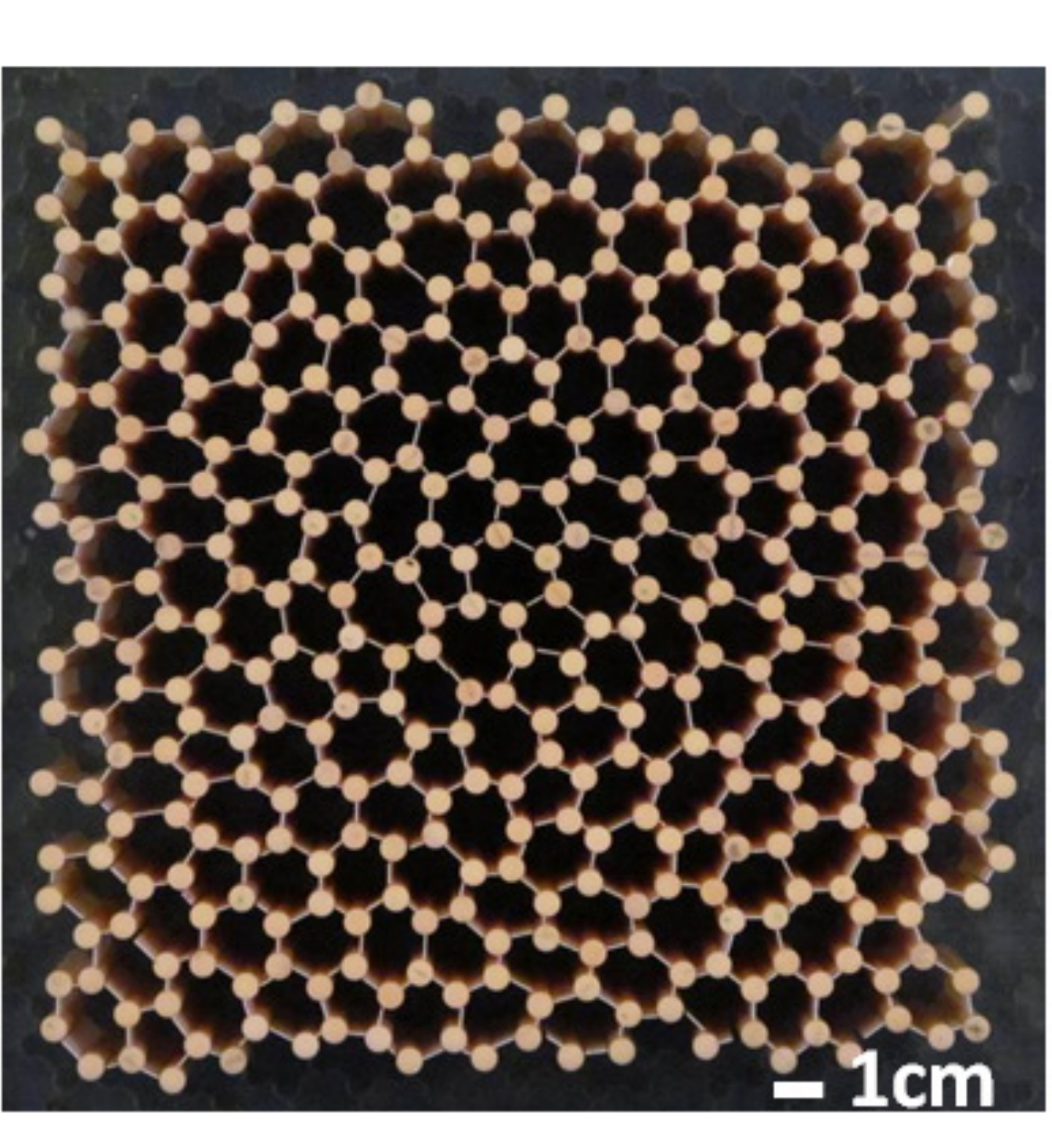}}}
\caption{ (a) Band structure for stealthy hyperuniform networks as a function of $\chi$, as predicted from the computational
study in Ref. \citenum{Fl09b}. From left to right,
$\chi=0.1,0.2,0.3,0.4$ and 0.5.
 The relative band-gap size, measured by
$\Delta \omega/\omega_C$ takes on the largest value of  10.26\% for the rightmost case of $\chi=0.5$.   
(b) 3D fabrication  of the computationally-designed maximal band-gap structure looking down from the top,
as adapted from Ref. \citenum{Man13b}. The solid phase is aluminum oxide.}
\label{PBG}
\end{figure*}

By mapping relatively large 2D disordered stealthy hyperuniform point configurations, obtained via the
collective-coordinate optimization procedure, \cite{Uc04b,Ba08} to certain 2D trivalent dielectric
networks via a Delaunay centroidal tessellation, \cite{Fl09b} what was thought to
be impossible at the time  became possible. Specifically, the first disordered
network solids to have large {\it complete} (both polarizations and blocking all directions) photonic band gaps comparable in
size to those in photonic crystals were identified, but with the additional advantage of perfect isotropy \cite{Fl09b}.
The computational designs consist of trivalent networks of cell walls with circular
cylinders at the nodes. The band structure was computed as a function of the degree of stealthiness $\chi$
(left panel of Fig. \ref{PBG}) and the case of $\chi$ nearly equal to 0.5 (with an accompanying substantial degree
of short-range order) leads to the maximal complete band-gap size in disordered hyperuniform dielectric networks.
This numerical investigation enabled the design and  fabrication of disordered cellular solids with the predicted
photonic band-gap characteristics for the microwave regime (right panel of Fig. \ref{PBG}), enabling unprecedented
free-form waveguide geometries unhindered by crystallinity and anisotropy, and robust to defects. \cite{Fl13,Man13b}
 Subsequently, stealthy hyperuniform materials were shown to have novel electromagnetic and elastic  wave propagation characteristics, including 
transparency to long-wavelength radiation,  \cite{Le16,Fr17,Ki20a,Ki20b,To21a} tunable diffusive and localization regimes, \cite{Fr17}
enhanced absorption of waves, \cite{Bi19} and singular  phononic band gaps. \cite{Gk17,Ro19,Roh20}

\begin{figure*}[bthp]
\centerline{\subfigure[]{\includegraphics[  width=3.in, keepaspectratio,clip=]{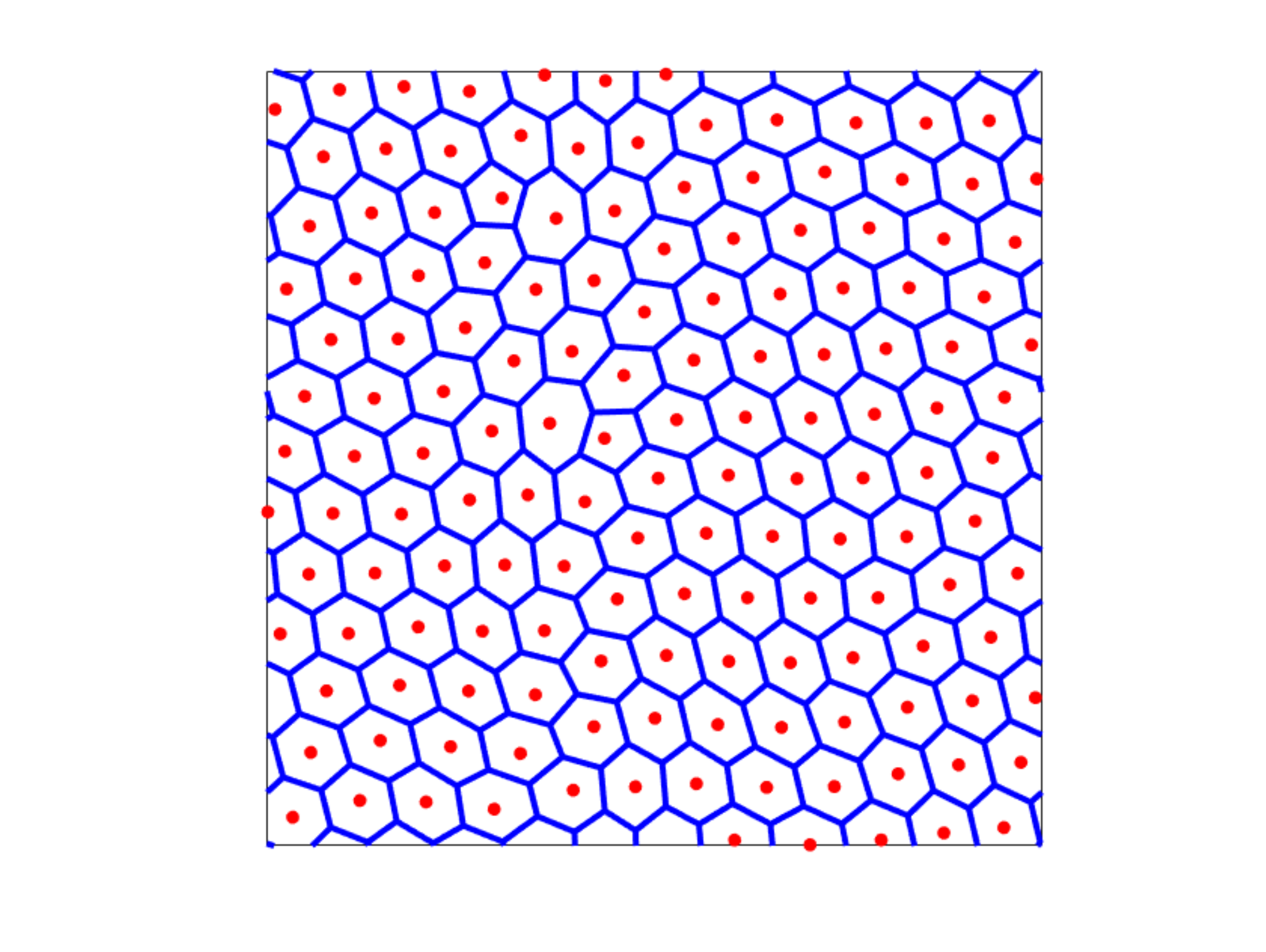}}\subfigure[]{\includegraphics[  width=2.6in, keepaspectratio,clip=]{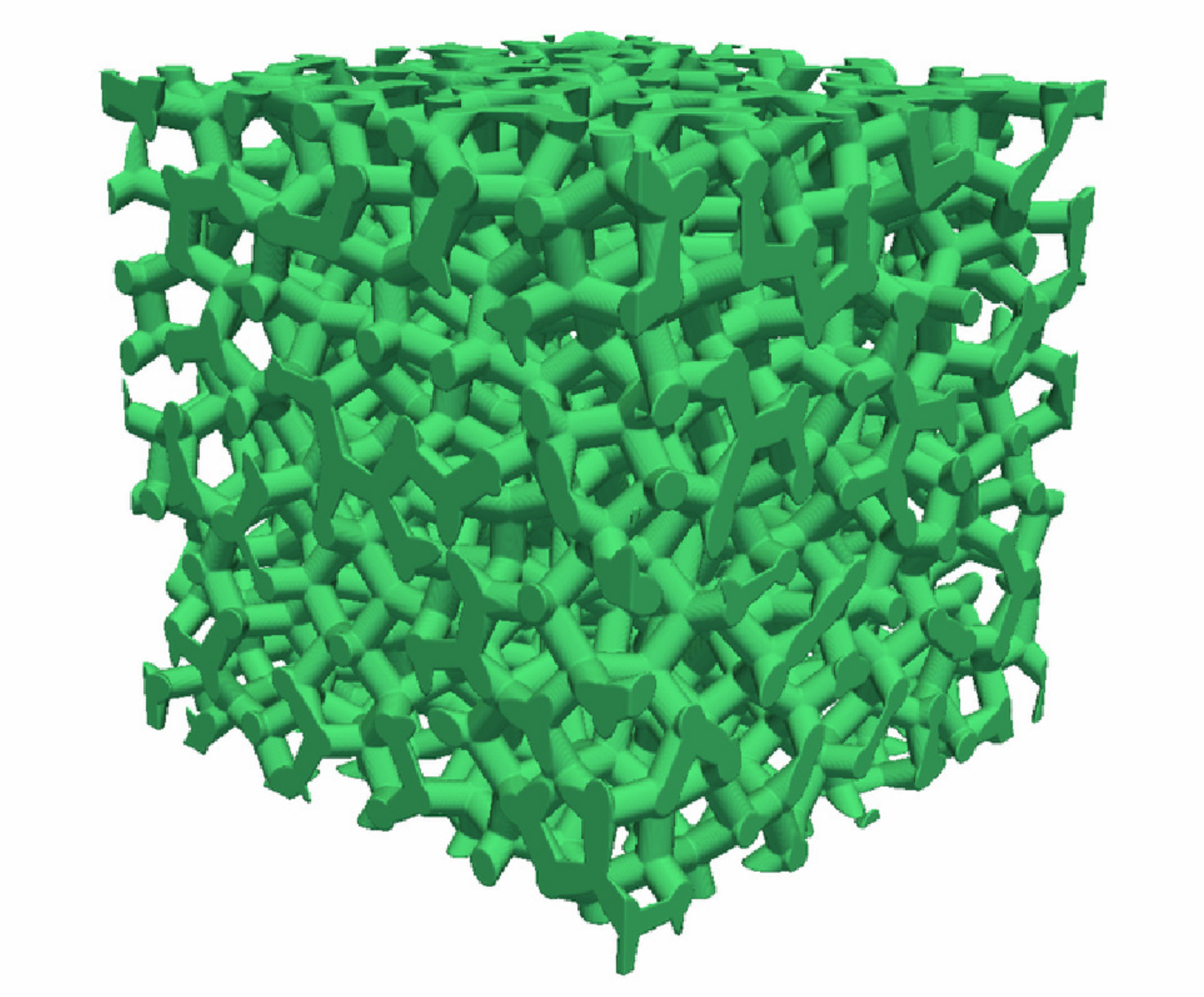}}}
\caption{(a) 2D disordered hyperuniform trivalent network, as adapted from Ref. \citenum{To18c}. (b) 3D disordered hyperuniform tetrahedrally-coordinated network,  as adapted from Ref. \citenum{Kl21a}.}
\label{networks}
\end{figure*}

The effective thermal or electrical conductivities and elastic moduli of 
various 2D ordered and disordered hyperuniform cellular networks were studied. \cite{To18c} The multifunctionality of a class of such low-density networks was established by demonstrating that they maximize or virtually maximize the effective conductivities and elastic moduli. This was accomplished by using the machinery of homogenization theory, including optimal
bounds and cross-property bounds, and statistical mechanics. It was rigorously proved that anisotropic
networks consisting of sets of intersecting parallel channels in the low-density limit, ordered or
disordered, possess optimal effective conductivity tensors. For a variety of different disordered
networks, it was shown that when short-range and long-range order increases, there is an increase in
both the effective conductivity and elastic moduli of the network. Moreover, it was demonstrated that
the effective conductivity and elastic moduli of various disordered networks (derived from
disordered ``stealthy" hyperuniform point patterns), such as the one shown in the left panel of Fig. \ref{networks}),  
possess virtually optimal values. Interestingly, the
optimal networks for conductivity are also optimal for the fluid permeability associated with slow
viscous flow through the channels as well as the mean survival time associated with diffusion-controlled
reactions in the channels. 3D disordered hyperuniform networks, such as the one shown in right panel
of Fig. \ref{networks}, have been shown to have sizable photonic band gaps. \cite{Kl21a} In summary, 2D and 3D disordered
hyperuniform low-weight cellular networks are  multifunctional with respect to transport
(e.g., heat dissipation and fluid transport), mechanical and electromagnetic properties, which can
be readily fabricated using  2D  lithographic and 3D printing technologies \cite{Wo12,Va13,Sh15,Zhao18}.

The theoretical problem of estimating the effective properties of multiphase composite media is an outstanding one and dates back to work by some of the luminaries of science, including Maxwell, \cite{Ma73} Lord Rayleigh, \cite{Ra92} and Einstein. \cite{Ei06}
The preponderance of previous  theoretical studies have focused on the determination of static effective properties (e.g., dielectric constant, elastic moduli and fluid permeability) using a variety of methods, including approximation schemes, \cite{Ma73,Br35,Br47,Bu65} bounding techniques, \cite{Pr61,Ha62c,Be65, Ko88b,To02a,Mi02} and exact series-expansion procedures. \cite{Br55,Fe82a, Se89,To97b}
%Even the prediction of the static effective dielectric constant $\boldsymbol{\varepsilon}_e$ is notoriously difficult
%for general dielectric  composite media because it depends on an infinite set of correlation functions that characterizes the microstructure.
Much less is  known about  the theoretical prediction of the effective dynamic dielectric constant tensor $\boldsymbol{\varepsilon}_e({\bf k}_I)$,
where ${\bf k}_I$ is wavevector of the incident radiation. The strong-contrast formalism has recently been used
to derive exact {\it nonlocal} expansions for $\boldsymbol{\varepsilon}_e({\bf k}_I)$ that exactly account for complete microstructural information  and hence multiple scattering
to all orders for the range of wavenumbers for which our extended homogenization theory applies, i.e.,  $0 \le  |{\bf k}_I| \ell \lesssim 1$
(where $\ell$ is a characteristic heterogeneity length scale).\cite{To21a}
Due to the fast-convergence properties of such expansions, their lower-order truncations yield accurate closed-form approximate formulas for  
$\boldsymbol{\varepsilon}_e({\bf k}_I)$ that depend on the spectral density ${\tilde \chi}_{_V}({\bf k})$. It was shown
that disordered stealthy hyperuniform particulate composites exhibit novel wave characteristics, including the capacity to act as low-pass filters that transmit waves
``isotropically” up to a selected wavenumber or refractive indices that abruptly change over a narrow range of wavenumbers.
The aforementioned nonlocal formulas can now  be used to accelerate the discovery of novel electromagnetic composites by appropriate tailoring of the spectral densities.

\begin{figure*}[bthp]
\centerline{\subfigure[]{\includegraphics[width=0.48\textwidth]{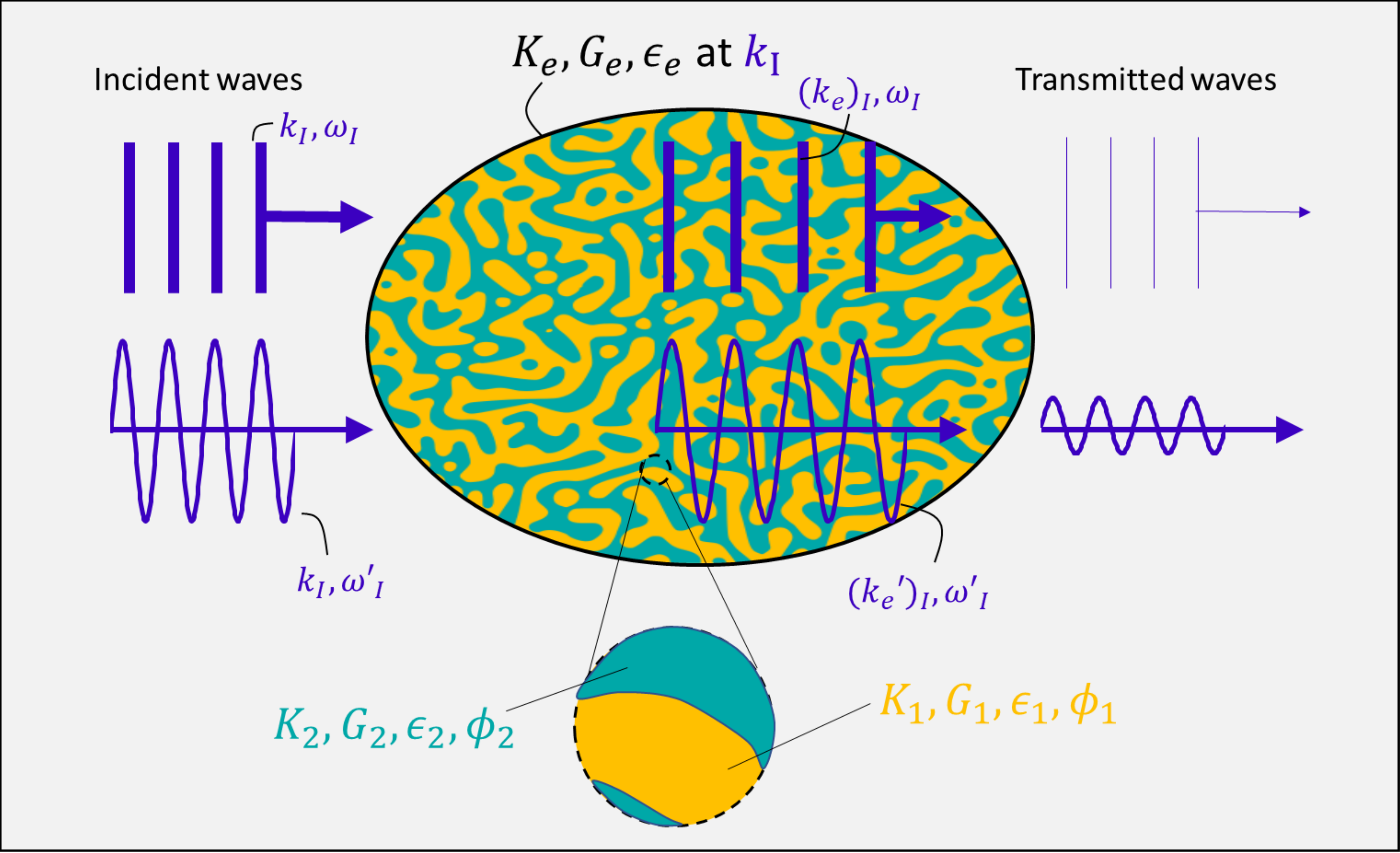}}\hspace{0.25in} 
            \subfigure[]{\includegraphics[width=0.48\textwidth]{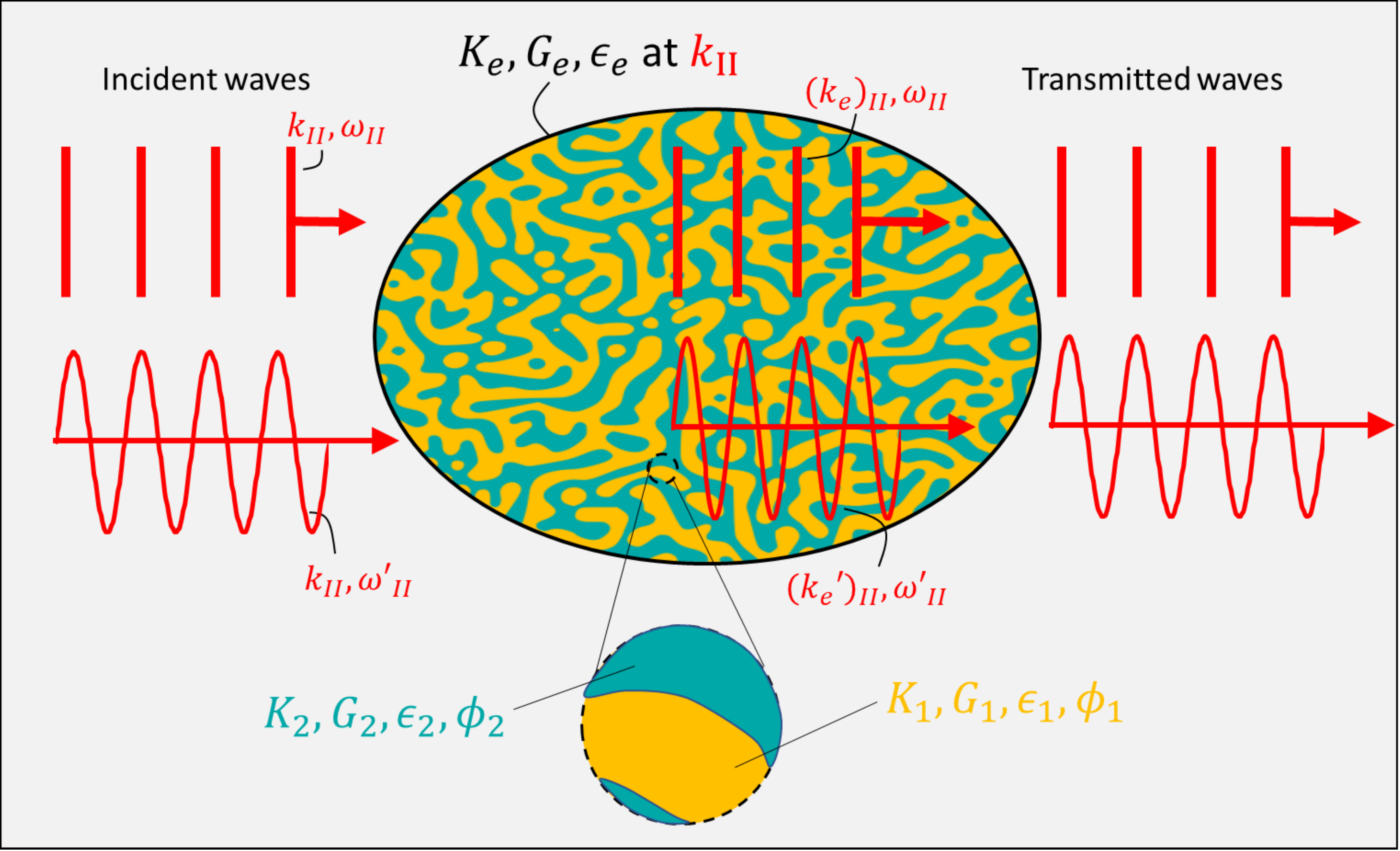}}}
\caption{  Schematics illustrating elastic and electromagnetic waves at two different wavenumbers (a) $k_I$ and (b) $k_{II}$  incident to, inside of and transmitted from a two-phase heterogeneous material (a large ellipse)  consisting of a matrix phase (shown in yellow) and a dispersed phase (shown in cyan).
 Parallel lines and sinusoidal curves represent elastic and electromagnetic waves, respectively.
(a) For an elastic wave  with a wavenumber $k_I$, while the wavefronts inside this material  experience microscopic disturbances, they effectively behave like a plane wave inside a homogeneous material with an effective wavenumber $(k_e)_{I}$ and effective elastic moduli $K_e$ and $G_e$.
 Analogously,  for an electromagnetic wave, this material behaves like a homogeneous material with an effective dielectric constant $\epsilon_e$.
For instance, both elastic and electromagnetic waves are attenuated due to scattering if this  composite has a non-zero scattering intensity at $k_I$.
 (b) For waves (red) of a wavenumber $k_{II}$, this composite can be effectively transparent, if it has a zero-scattering intensity at $k_{II}$. This
figure is adapted from Ref. \citenum{Ki20a}.
\label{fig:schem}}
\end{figure*}

Cross-property relations for two-phase composite media were recently obtained that link effective elastic and electromagnetic wave 
characteristics to one another, including effective wave speeds and attenuation coefficients.\cite{Ki20a} This was achieved by deriving 
accurate formulas for the effective elastodynamic properties  \cite{Ki20b} as well as effective electromagnetic properties,\cite{To21a}
each of which depend on the microstructure via the spectral density. Such formulas enable one to explore
the wave characteristics of a broad class of disordered microstructures, including exotic disordered hyperuniform
varieties.  It was  specifically demonstrated that disordered stealthy hyperuniform/nonhyperuniform
microstructures exhibit novel elastic wave characteristics that have the potential for future applications, e.g.,
narrow-band or narrow-band-pass filters that absorb or transmit elastic waves isotropically for a narrow 
spectrum of frequencies, respectively. These cross-property relations for effective electromagnetic and elastic wave characteristics can be applied to
design multifunctional composites (Fig. \ref{fig:schem}), such as exterior components of spacecrafts or building materials that require
both excellent stiffness and electromagnetic absorption, and heat-sinks for CPUs that have to efficiently emit
thermal radiation and suppress mechanical vibrations, and nondestructive evaluation of the mechanical strength
of materials from the effective dielectric response.

%\begin{figure}[H]
%\begin{center}

%\includegraphics[width=2.5in,clip=,keepaspectratio]{/u/torquato/Papers/Kim/Waves/Part1/fig9b-modified}
%\end{center}
%\end{figure}

\begin{figure*}[bthp]
\subfigure[]{
\includegraphics[  width=2.in, keepaspectratio,clip=]{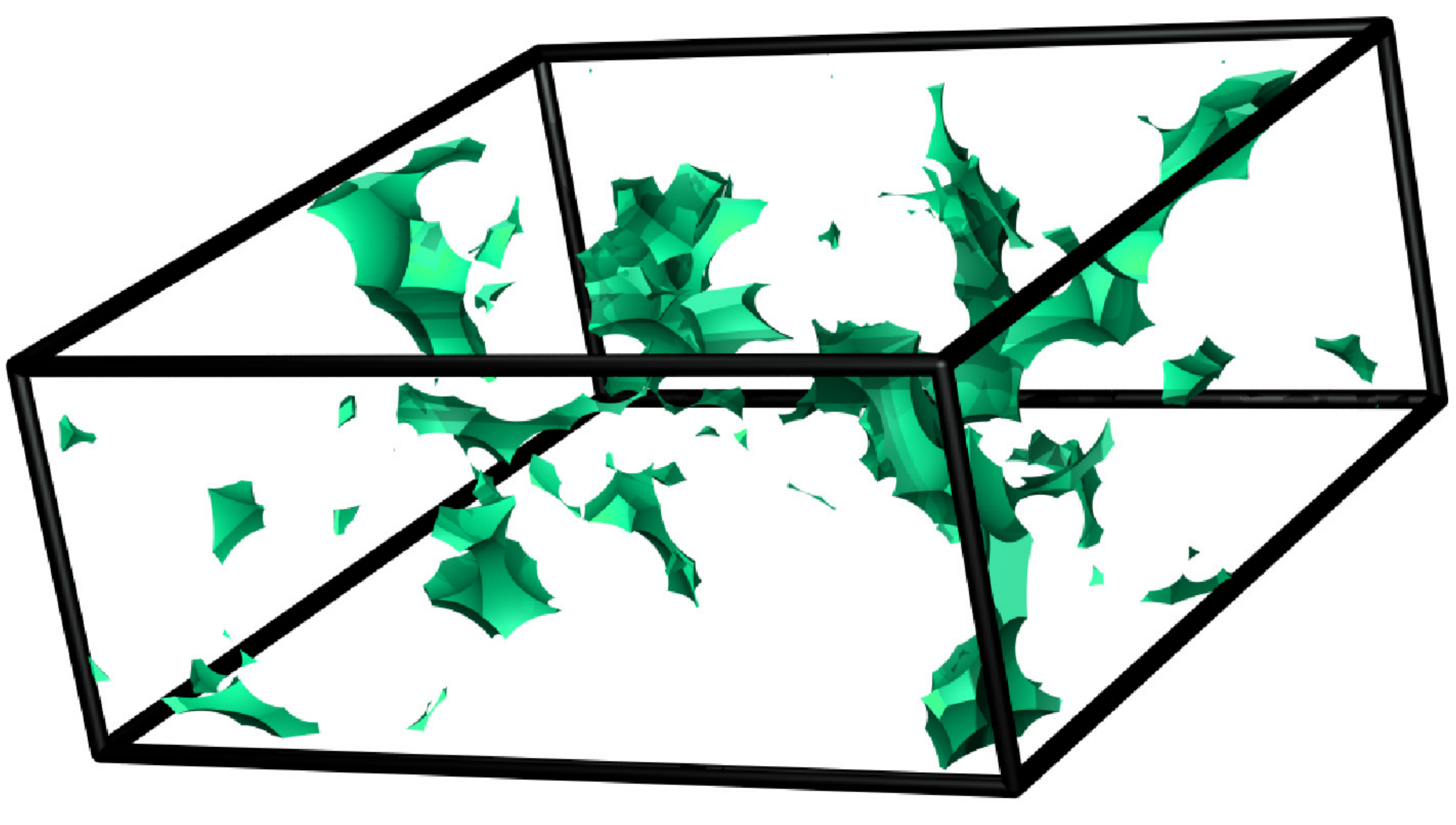}}
\subfigure[]{
\includegraphics[  width=2.in, keepaspectratio,clip=]{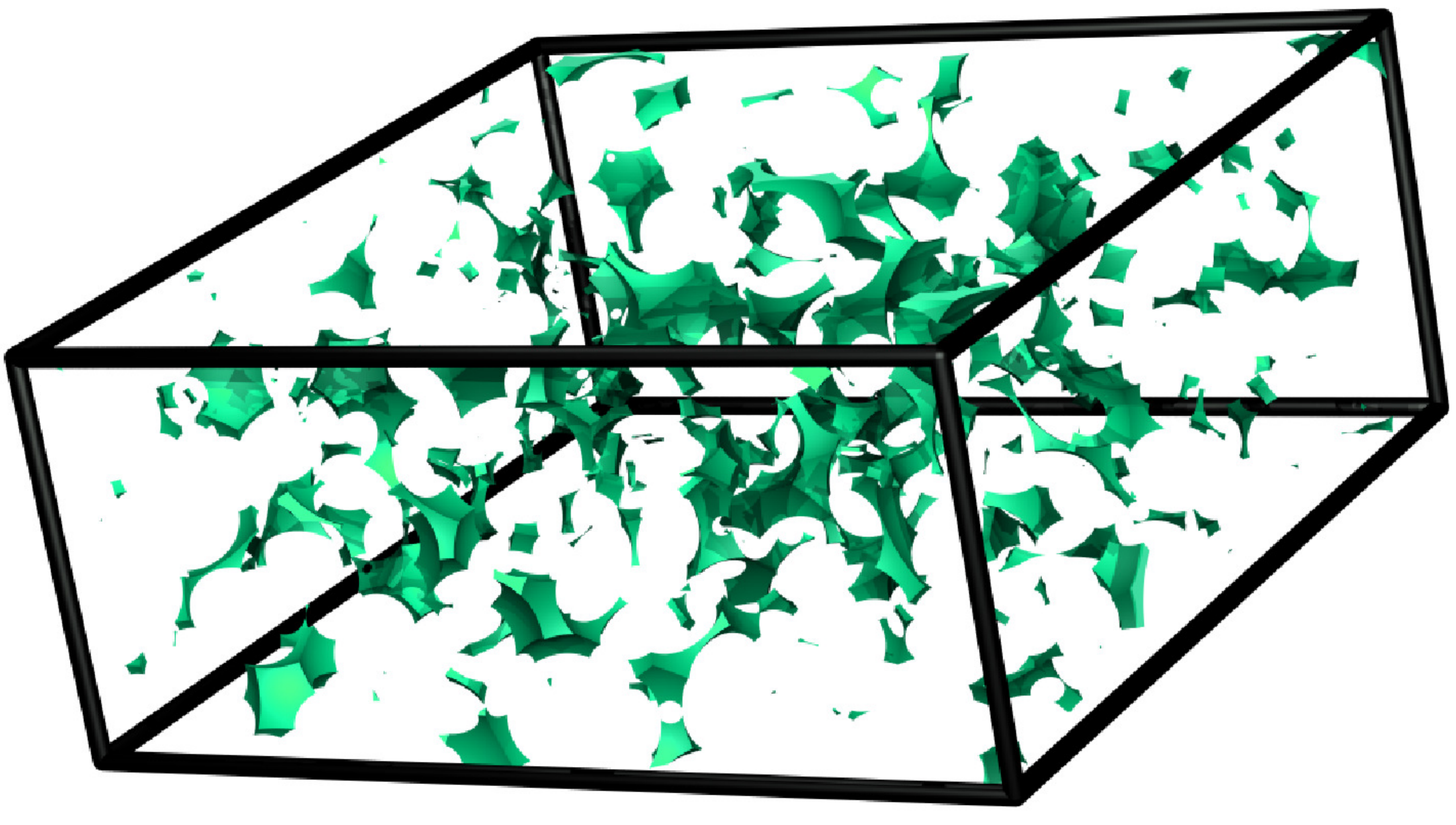}}
\subfigure[]{
\includegraphics[  width=2.in, keepaspectratio,clip=]{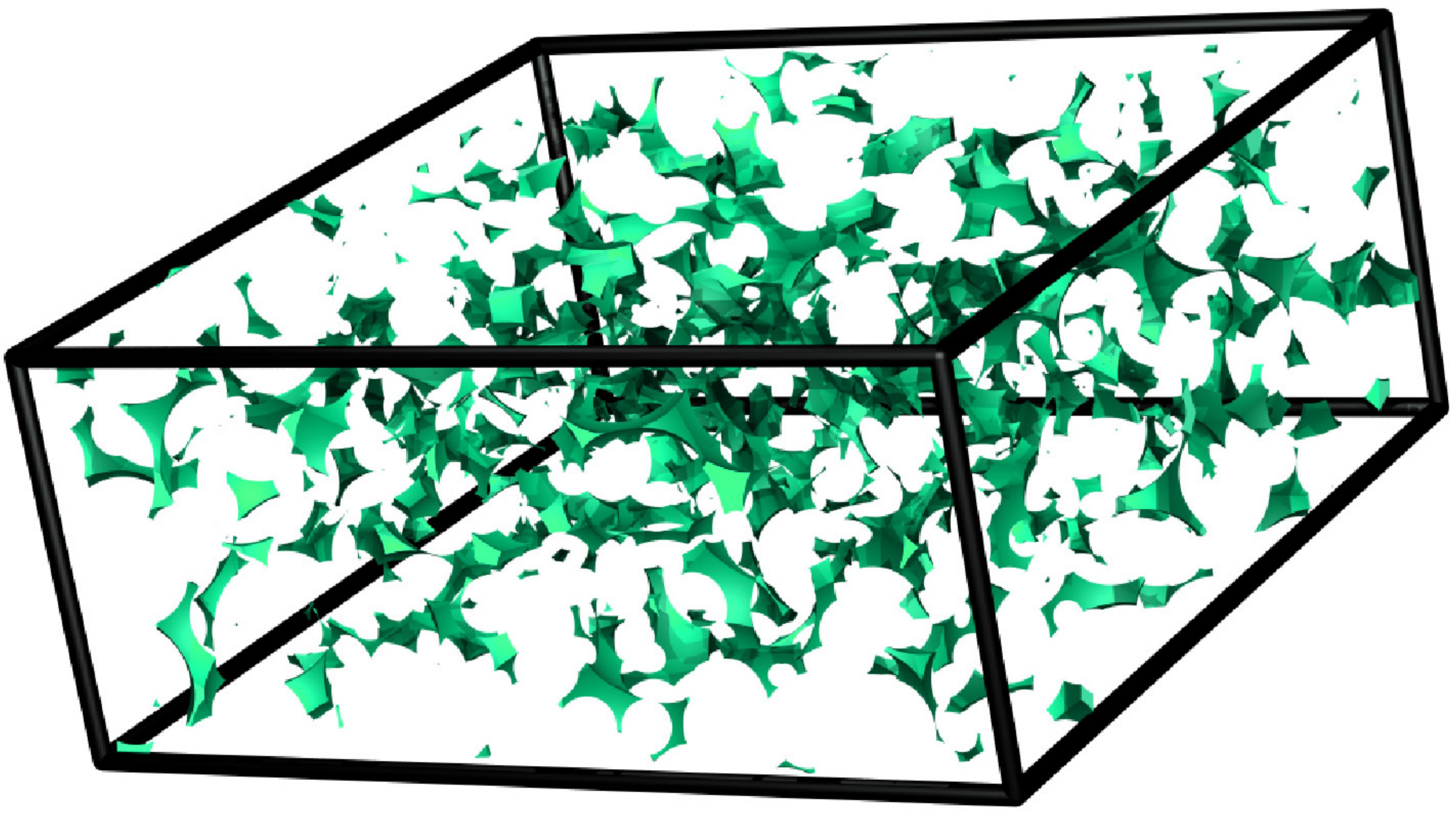}}
\caption{Images of the void space after applying a ``dilation" operation to three different sphere packings, as adapted from Ref. \citenum{To20a}. 
(a) A nonhyperuniform equilibrium packing. (b) A hyperuniform MRJ packing. (c)
A disordered stealthy packing. }
\label{dilation}
\end{figure*}

The effective transport characteristics of fluid-saturated porous media have been studied
using certain rigorous microstructure-property relations. \cite{To20a} Of particular interest were the predictions of the formation factor $\cal F$,
mean survival time $\tau$, principal NMR (diffusion) relaxation time $T_1$,
principal viscous relaxation time $\Theta_1$, and fluid permeability $k$ for hyperuniform and nonhyperuniform models of porous media.
 Among other results, 
a Fourier representation of a classic rigorous upper bound on the fluid permeability was derived
that depends on the spectral density ${\tilde \chi}_{_V}({\bf k})$  to infer  how the permeabilities of hyperuniform porous media perform relative to those of nonhyperuniform ones; see Fig. \ref{dilation}. It was found  that the velocity
fields in nonhyperuniform porous media are generally much more localized over the pore space
compared to those in their  hyperuniform counterparts, which has certain implications for their permeabilities.
Rigorous bounds on the transport properties $\cal F$, $\tau$, $T_1$ and $\Theta_1$ suggest a new approximate formula for the
fluid permeability that provides reasonably accurate permeability predictions of a certain class of hyperuniform and nonhyperuniform porous media. These comparative studies  shed new light on the microstructural characteristics, such as {\it pore-size statistics}, in determining the transport
properties of general porous media. In a more recent study,  the second moment of the pore-size probability density function
was shown to be correlated with the {\it critical pore radius}, which contains crucial {\it connectivity}
information about the pore space. \cite{Kl21b}
All of these findings  have important implications for the design of porous materials with desirable transport properties.

\begin{figure}[bthp]
\centerline{\includegraphics[width=3.3in,keepaspectratio,clip=]{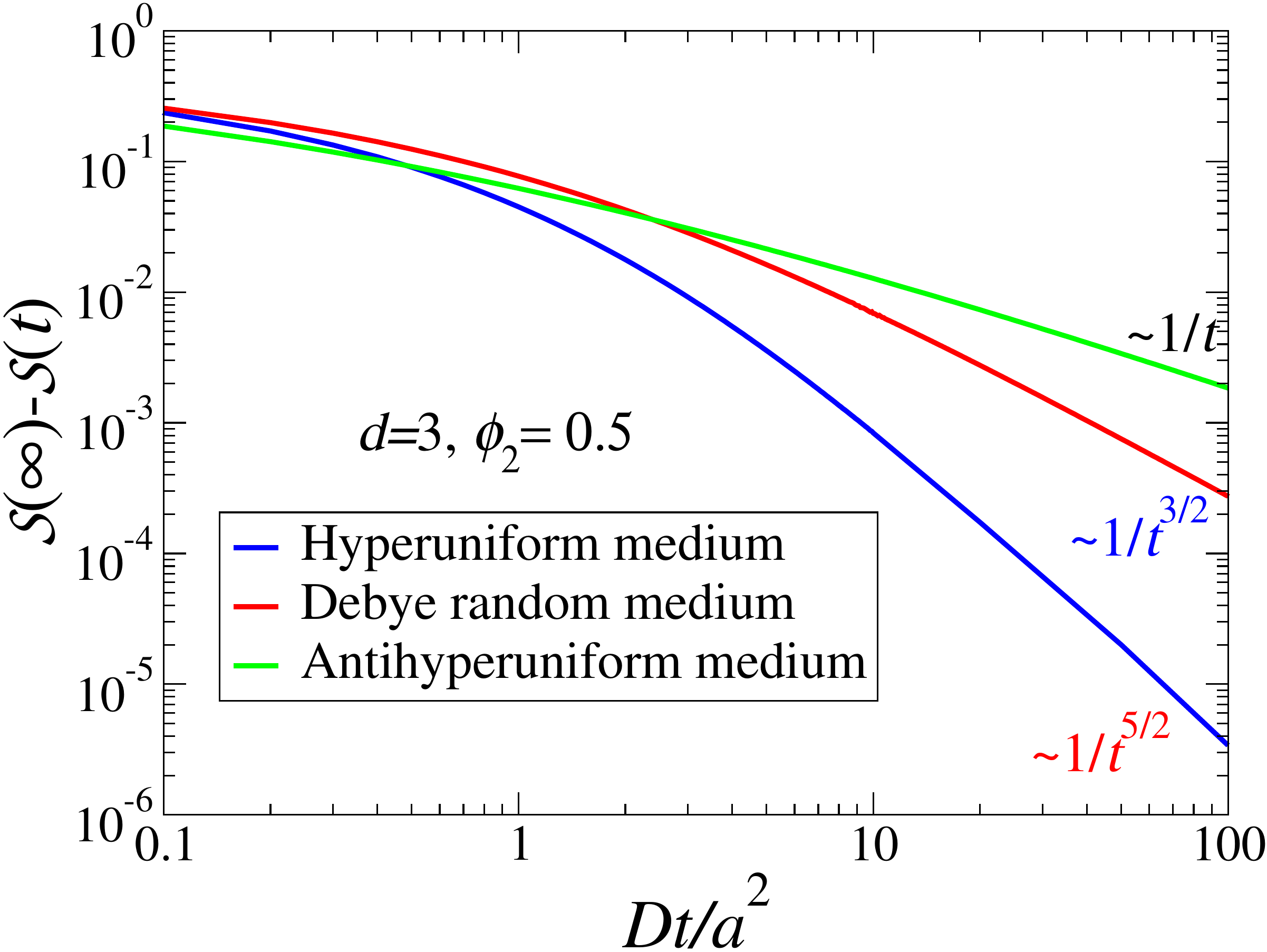}}
\caption{Excess spreadabilities versus dimensionless time $D t/a^2$ for antihyperuniform media (top curve), Debye random media (middle curve),
and disordered hyperuniform media (bottom curve) for $d = 3$ and
$\phi_2 = 0.5$, as adapted from Ref. \citenum{To21d}. The long-time inverse power-law scalings of ${\cal S}(\infty) -{\cal S}(t)$
for each of these models is indicated. Here $a$ is a characteristic length scale for each model, as defined in Ref. \citenum{To21d}. }
\label{spread}
\end{figure}

A new dynamic probe of the microstructure
of two-phase media has been introduced called the {\it spreadability} ${\cal S}(t)$, which is
a  measure of the spreadability of diffusion information
as a function of time $t$ in any Euclidean space dimension $d$. \cite{To21d}
It is assumed that a solute at $t=0$ is uniformly distributed
throughout phase 2 with volume fraction $\phi_2$, and completely absent from
phase 1 with volume fraction $\phi_1$, and each phase has same diffusion
coefficient $D$. The spreadability  is the fraction of the
total amount of solute present that has diffused into
phase 1 at time $t$.
In particular, a three-dimensional formula due to Prager \cite{Pr63a}
was generalized to any dimension in direct space and its Fourier representation
was derived. The latter is an exact integral relation for the spreadability ${\cal S}(t)$ that depends
only on the  spectral density ${\tilde \chi}_{_V}({\bf k})$:
\begin{equation}
{\cal S}(\infty)- {\cal S}(t)=\frac{1}{(2\pi)^d\,\phi_2} \int_{\mathbb{R}^d} {\tilde \chi}_{_V}({\bf k}) \exp[-k^2 Dt] d{\bf k} \ge 0,
\label{4}
\end{equation}
where ${\cal S}(\infty)=\phi_1$.
Importantly, the short-, intermediate- and long-time behaviors of ${\cal S}(t)$ contain crucial
small-, intermediate- and large-scale structural characteristics.
For hyperuniform media, it was shown that the ``excess” spreadability, ${\cal S}(\infty)-{\cal S}(t)$, decays to its
long-time behavior exponentially faster than that of any nonhyperuniform medium, the ``slowest" being
antihyperuniform media, as illustrated in Fig. \ref{spread}.
It was  also shown that there is a remarkable link between the spreadability and  nuclear magnetic resonance (NMR) pulsed field
gradient spin-echo  amplitude and diffusion MRI. \cite{To21d} Elsewhere,
this new theoretical/experimental tool was applied to characterize many different models
and a porous-medium sample. \cite{Wa22a,Ma22a}

\section{Conclusions and Outlook}
\label{conclusions}

We have seen that the exotic hybrid crystal-liquid structural attributes of disordered hyperuniform composites
can be endowed with an array of extraordinary physical properties, including photonic, phononic, transport and mechanical characteristics 
that are only beginning to be discovered. Disordered hyperuniform media can have advantages over their periodic counterparts, 
such as unique or nearly optimal, direction-independent
physical properties and robustness against defects. The field of hyperuniformity is still in its infancy, though,
and a deeper fundamental understanding of these unusual states of matter is required
in order to realize their full potential for next-generation materials.
Future challenges include the further development of forward and inverse computational approaches to generate
disordered hyperuniform structures, formulation of    improved order metrics to rank order them,
and identifying their desirable multifunctional characteristics.
These computational designs can subsequently be
combined with the  2D lithographic fabrication techniques \cite{Zhao18} and 3D additive manufacturing techniques \cite{Wo12,Va13,Sh15}
to accelerate the discovery of novel multifunctional hyperuniform two-phase materials.

Cross-property ``maps" have recently been introduced to connect combinations of pairs of 
effective static transport and elastic properties of general particulate media via
analytical structure-property formulas. \cite{To18e} Cross-property maps and
their extensions will facilitate the rational design of composites with different desirable multifunctional
characteristics. In future work, it would be valuable to formulate cross-property
maps for the various physical properties described in Sec. \ref{novel-multi} using the corresponding analytical estimates of 
these properties in order to aid in the multifunctional design of disordered hyperuniform composites.

To complement rigorous approaches to estimate the macroscopic properties of heterogeneous media from the microstructure, 
data-driven methodologies to establish structure-property relationships are increasingly being
employed.\cite{Va16,Li18,Ne20,Ro20} The rapid increase in computational resources facilitates the calculation of effective properties for very large data sets 
(thousands or more) of different microstructures, including those obtained experimentally 
via 2D and 3D  high-resolution imaging techniques. \cite{Co96,Na01,To02a,Bl13,Re22}  As a result, it has  become manageable to generate large numbers of realistic 
virtual microstructures, and using those to perform exploratory computational
screening of structure–property relationships. The application of machine-learning and other
data-driven approaches for the discovery of multifunctional disordered hyperuniform composites
has yet to be undertaken and hence is a promising avenue for future research.

\begin{acks}
The author thanks Jaeuk Kim, Charles Maher, Murray Skolnick and Haina Wang
for the valuable comments about the paper. The author gratefully acknowledges the
support of the Air Force Office of Scientific Research Program on
Mechanics of Multifunctional Materials and Microsystems under
award No. FA9550-18-1-0514.
\end{acks}

%\bibliographystyle{SageV}
%\bibliography{new}

\end{document}